\documentclass[useAMS,usenatbib]{mn2e}
\usepackage{epsfig,color,pifont}
\usepackage{rotating}

\usepackage{lscape}

\usepackage{natbib}
\bibpunct{(}{)}{;}{a}{}{,}
\def\spose#1{\hbox to 0pt{#1\hss}}
\def\ltsimm{\mathrel{\spose{\lower 3pt\hbox{$\sim$}}
        \raise 2.0pt\hbox{$<$}}}
\def\gtsimm{\mathrel{\spose{\lower 3pt\hbox{$\sim$}}
        \raise 2.0pt\hbox{$>$}}}

\def\km{{\rm\thinspace km}}
\def\cm{{\rm\thinspace cm}}

\def\s{{\rm\thinspace s}}
\def\yr{{\rm\thinspace yr}}
\def\g{{\rm\thinspace g}}
\def\kmps{\hbox{${\rm\km\s^{-1}\,}$}}

\def\erg{{\rm\thinspace erg}}
\def\Hz{{\rm\thinspace Hz}}

\def\ster{{\rm\thinspace ster}}
\def\ergps{\hbox{${\rm\erg\s^{-1}\,}$}}

\def\Msol{\hbox{${\rm\thinspace M_{\odot}}$}}

\def\Msolpyr{\hbox{${\rm\Msol\yr^{-1}\,}$}}
\def\pcm{\hbox{${\rm\cm^{-1}\,}$}}
\def\pcm2{\hbox{${\rm\cm^{-2}\,}$}}
\def\pcm3{\hbox{${\rm\cm^{-3}\,}$}}
\def\ergpscm3Hz{\hbox{${\rm\ergps\cm^{-3}\Hz^{-1}\,}$}}
\def\ergpscm3Hzster{\hbox{${\rm\ergps\cm^{-3}\Hz^{-1}\ster^{-1}\,}$}}
\def\gpcm3{\hbox{${\rm\g\cm^{-3}\,}$}}
\def\ergpcm2{\hbox{${\rm\erg\cm^{-2}\,}$}}
\def\ergpcm3{\hbox{${\rm\erg\cm^{-3}\,}$}}
\def\phpscm2{\hbox{${\rm photons\s^{-1}\cm^{-2}\,}$}}

\def\aap{{\rm A\&A}}
\def\apj{{\rm ApJ}}
\def\apjs{{\rm ApJS}}
\def\aj{{\rm AJ}}
\def\mnras{{\rm MNRAS}}
\def\nat{{\rm Nature}}

\def\araa{{\rm ARA\&A}}

\def\pasp{{\rm PASP}}
\def\pasj{{\rm PASJ}}

\title[The interactions of winds from massive YSOs]{The interactions
of winds from massive young stellar objects: X-ray emission, dynamics,
and cavity evolution}

\author[E.~R.~Parkin, J.~M.~Pittard, M.~G.~Hoare, N.~J.~Wright,
J.~J.~Drake] {E. R. Parkin$^{1}$\thanks{E-mail: erp@ast.leeds.ac.uk},
J.~M.~Pittard$^{1}$\thanks{E-mail: jmp@ast.leeds.ac.uk},
M.~G.~Hoare$^{1}$\thanks{E-mail: mgh@ast.leeds.ac.uk},
N.~J.~Wright$^{2}$\thanks{E-mail: nwright@cfa.harvard.edu},
J.~J.~Drake$^{2}$\thanks{E-mail: jdrake@cfa.harvard.edu} \\
$^{1}$School of Physics and Astronomy, The University of Leeds,
Woodhouse Lane, Leeds LS2 9JT, UK \\ $^{2}$Harvard-Smithsonian Center
for Astrophysics, 60 Garden Street, Cambridge MA01238, USA}

\begin{document}

\date{Accepted ... Received ...; in original form ...}

\pagerange{\pageref{firstpage}--\pageref{lastpage}} \pubyear{2009}

\maketitle

\label{firstpage}

\begin{abstract}
  2D axis-symmetric hydrodynamical simulations are presented which
  explore the interaction of stellar and disk winds with surrounding
  infalling cloud material. The star, and its accompanying disk, blow
  winds inside a cavity cleared out by an earlier jet. The collision
  of the winds with their surroundings generates shock heated plasma
  which reaches temperatures up to $\sim
  10^{8}\thinspace$K. Attenuated X-ray spectra are calculated from
  solving the equation of radiative transfer along
  lines-of-sight. This process is repeated at various epochs
  throughout the simulations to examine the evolution of the intrinsic
  and attenuated flux. We find that the dynamic nature of the
  wind-cavity interaction fuels intrinsic variability in the observed
  emission on timescales of several hundred years. This is principally
  due to variations in the position of the reverse shock which is
  influenced by changes in the shape of the cavity wall. The collision
  of the winds with the cavity wall can cause clumps of cloud material
  to be stripped away. Mixing of these clumps into the winds
  mass-loads the flow and enhances the X-ray emission measure. The
  position and shape of the reverse shock plays a key role in
  determining the strength and hardness of the X-ray emission. In some
  models the reverse shock is oblique to much of the stellar and disk
  outflows, whereas in others it is closely normal over a wide range
  of polar angles. For reasonable stellar and disk wind parameters the
  integrated count rate and spatial extent of the intensity peak for
  X-ray emission agree with \textit{Chandra} observations of the
  deeply embedded MYSOs S106 IRS4, Mon R2 IRS3 A, and AFGL 2591.

The evolution of the cavity is heavily dependent on the ratio of the
inflow and outflow ram pressures. The cavity closes up if the inflow
is too strong, and rapidly widens if the outflowing winds are too
strong. The velocity shear between the respective flows creates
Kelvin-Helmholtz (KH) instabilities which corrugate the surface of the
cavity. Rayleigh-Taylor-like instabilities also occur when the cavity
wall is pushed forcefully backwards by strong outflows. The opening
angle of the cavity plays a significant role and we find that for
collimation factors in agreement with those observed for bi-polar jets
around massive young stellar objects (MYSOs), a reverse shock is
established within $\ltsimm 500\;{\rm au}$ of the star.

\end{abstract}


\section{Introduction}
\label{sec:intro}

Observational and theoretical advances have provided increasing
evidence that massive star formation is not merely a scaled up version
of low mass star formation. The crucial difference arises from the
prominent role of radiation pressure in massive star formation
\citep[][and references there-in]{Zinnecker:2007}. Theoretical models
find that the radiation pressure from the core is too great for
spherical accretion to form stars above a mass of $\sim 10 \Msol$
\citep{Wolfire:1987}. This problem disappears if one considers disk
accretion \citep{Yorke:2002,Krumholz:2009}. Observations of young
massive stars support this scenario as they show the presence of a
dense equatorial disk and ongoing accretion
\citep{Patel:2005,Beltran:2006,Torrelles:2007}. There are, however,
some similarities in the formation processes of both low and high mass
stars. For instance, both involve outflows \citep{Garay:1999,
Reipurth:2001, Banerjee+Pudritz:2006, Banerjee:2007}. Bi-polar
cavities around YSOs are commonly observed in star formation
\citep{Garay:1999}, with collimation factors of $\sim2-10$ for MYSOs
\citep{Beuther:2002b, Davis:2004}. The cavities may be more strongly
collimated at large distances (due to their formation by jets), but
less so at the base due to the action of stellar winds
\citep{Shepherd:2005}. For low-mass stars the observed cavity
morphology can be qualitatively reproduced by the injection of an
outflow into a non-spherical density distribution
\citep{Delamarter:2000, Lee:2001, Wilkin:2003, Mendoza:2004,
Cunningham:2005}. Magnetic fields increase the collimation
\citep{Gardiner:2003,Shang:2006}. Numerical simulations of high-mass
star formation show that outflows are a by-product of the collapse
process which forms the massive star, and may be due to radiation
pressure and magnetic fields \citep{Yorke:2002,Banerjee:2007}.

The widths of IR recombination line emission observed from MYSOs
indicates the presence of dense outflows with velocities ranging from
100 to $>340\;{\rm km\thinspace s}^{-1}$
\citep{Drew:1993,Bunn:1995}. Further confirmation of outflows has come
from high angular resolution radio observations
\citep[e.g.][]{Hoare:1994,Hoare:2006}. One explanation would be a disk
wind generated when UV flux from the star is absorbed and re-emitted
by the material at the surface of the disk.

Radiation-driven accretion disk winds have been previously explored
using hydrodynamic models. \cite{Proga:1998} and \cite{Drew:1998}
found that for a star-disk system where the $10\Msol$ star and it's
accompanying disk were equally luminous (i.e. $L_{\ast}/L_{\rm disk} =
1$), a strong steady disk wind could be produced. They found that
there was a transition from a fast, low density polar outflow to a
higher density outflow at a polar angle of $\theta \simeq
40-60^{\circ}$.  The stellar wind had a velocity of $\simeq 2000
\;{\rm km\thinspace s^{-1}}$, and the disk wind velocity ranged from
$\simeq 400\;\kmps$ at high latitudes to tens of \kmps at low
latitudes. They also found a density contrast of $10^{2}-10^{4}$
between the stellar and disk winds, the latter being much denser. The
opening angles of the stellar and disk winds are dependent on a number
of parameters in the disk wind model: the luminosity ratio between the
star and the disk, the force parameter used to calculate the radiative
driving \citep{Proga:1999}, and the magnetic pressure in the wind
\citep{Proga:2003}. However, the assumption of strong magnetic fields
in MYSO disks is speculative. The equatorial flow detected around S106
IRS4 (or IR) shows a mass-loss rate of $\dot{M} \simeq 10^{-6}
\Msolpyr$ \citep{Drew:1993,Hoare:1994}, a factor of $\sim100$ higher
than the simulations. Such a difference could be accounted for by
clumping in the wind. \cite{Sim:2005} used a radiative transfer model
inspired by disk wind simulations to model the HI line emission from
MYSOs. They adopted a disk wind which was largely equatorial and had a
velocity which declined from $\sim 850$ to $300\kmps$ in the angular
range $82-89^{\circ}$. This model proved to be successful in
reproducing qualitative features of Brackett and Pfund series line
emission observations.

X-rays have been detected from deeply embedded MYSOs in star forming
regions by the \textit{Chandra X-ray observatory} (hereafter
\textit{Chandra})\citep[e.g.][]{Broos:2007,Wang:2007,Wang:2009}. An
initial observation of Mon R2 detected X-ray emission from the
intermediate mass objects consistent with the hard and highly time
variable X-ray emission caused by magnetic flaring activity between
the star/disk \citep{Kohno:2002}. Further analysis of the same data,
coupled with high resolution near-IR interferometry identified the
separate constituents of Mon R2 IRS3. \cite{Preibisch:2002} found that
the X-ray emission from IRS3 A and C, with a count rate of $0.166\pm
0.041\;{\rm ks^{-1}}$ for the former, could not be explained by the
standard scenario for massive stars \citep[i.e. wind embedded shocks
produced by instabilities inherent in radiatively-driven winds -
see][]{Owocki:1988}, yet the estimated stellar masses of these objects
implies they will have radiative outer envelopes which poses problems
for the generation of X-rays through magnetic star/disk
interactions. In their observations of the S106 region,
\cite{Giardino:2004} suggested that it may not be necessary to employ
magnetic fields to explain the X-ray emission characteristics of the
massive central object S106 IRS4 where the count rate was
$0.30\pm0.11\;{\rm ks^{-1}}$.

A potential source of X-rays which has not been considered before is
the collision between the stellar and disk winds and the infalling
envelope. To explore this interaction we perform the first detailed
modelling of an MYSO embedded at the centre of a bi-polar cavity. The
star and disk drive supersonic outflows, and we examine how these
outflows influence the evolution of the infalling envelope. We find
that the wind-cavity interaction produces a reverse shock where gas is
heated to X-ray emitting temperatures. Our models are able to
reproduce the observed count rates, although there is a strong
sensitivity to the inclination angle of the observer.

This paper is organised as follows: \S~\ref{sec:model} gives a
description of the model used in the simulations. The results of the
simulations and the observable X-ray emission are presented in
\S~\ref{sec:results}. In \S~\ref{sec:fitstoobjects} we report on a
recent observation of the MYSO AFGL 2591 with \textit{Chandra} then
perform fits to this observation and those of two other candidate
objects, \S~\ref{sec:discussion} includes discussion, and in
\S~\ref{sec:conclusions} we summarize the conclusions from the current
work and suggest possible future directions.

\vspace{-5mm}
\section{The Model}
\label{sec:model}

\begin{figure}
\begin{center}
    \begin{tabular}{c}
      \resizebox{80mm}{!}{{\includegraphics{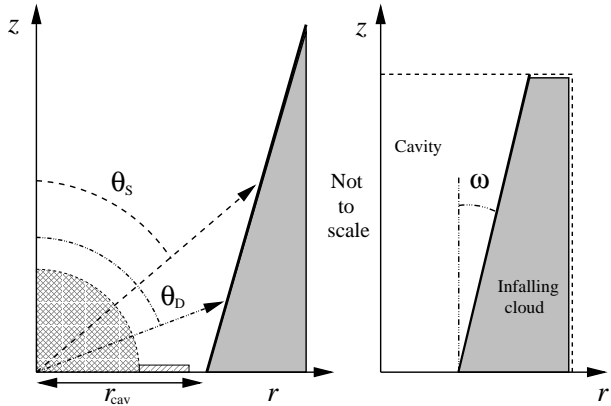}}} \\
    \end{tabular}
    \caption[]{Schematic of the simulation domain. The left-hand panel
      shows a zoom in of the region close to the star and disk. The
      right panel shows the entire simulation domain, with dotted-line
      representing the edge of the grid. The cross-hatched quadrant in
      the left-hand panel represents the region where the winds are
      mapped onto the grid and the adjacent rectangular region
      represents the additional cells used to ensure the angular
      dependence of the wind is resolved (see
      \S~\ref{subsec:winds}).}
    \label{fig:schematic}
\end{center}
\end{figure}

Our model is of a recently formed massive star with an accompanying
accretion disk. The star and disk are situated at the centre of a
flattened envelope with a pre-existing bi-polar cavity evacuated by
earlier jet activity. The morphology of the cavity is prescribed by a
simple analytic equation. A schematic of the simulation domain is shown in
Fig.~\ref{fig:schematic}.

The cavity intersects the disk plane at the cavity radius, $r_{\rm
  cav}$. In essence, the presence of a disk is implied through the
presence of the disk wind, yet we do not include any parameters for
the disk itself. We simply note that the disk wind is launched from
the inner regions of the disk \citep[i.e. $r\ltsimm 10\;{\rm
    R_{\ast}}$,][]{Drew:1998}. The star and disk are situated in the
first cell within the simulation domain. Initially the cavity is
filled with unshocked stellar and disk wind material, with the density
and velocity a function of the polar angle, $\theta$. The material
outside of the cavity is infalling, and represents molecular gas which
is still accreting onto the disk.

Features of the model, details of the hydrodynamics code used, and a
description of the X-ray calculations are now given.

\subsection{The infalling cloud}

For the cloud material we use the solutions for the collapse of a
slowly rotating isothermal sphere with a single point source of
gravity at the centre
\citep{Ulrich:1976,Cassen:1981,Chevalier:1983,Terebey:1984}. With
conservation of angular momentum for infalling material the
trajectories are parabolic and are described by the streamline
equation (in polar coordinates),
\begin{equation}
  \zeta = \frac{r_{\rm c}}{R} = \frac{\mu_{0} - \mu}{(1 - \mu_{0}^{2})\mu_{0}} \\
\label{eqn:streamline}
\end{equation}

\noindent where $\mu = \cos\theta$, $\mu_{0} = \cos\theta_{0}$, and $R
= \sqrt{z^{2} + r^{2}}$. $r_{\rm c}$ is the centrifugal radius of
the cloud and $\theta_{0}$ is the initial polar angle. The components
of the infall velocity are given by,

\begin{equation}
\begin{array}{lll}
  & u_{\rm R} = & -(\frac{G M_{\ast}}{R})^{1/2}(1 + \frac{\mu}{\mu_{0}}), \\
  & u_{\theta} = & (\frac{G M_{\ast}}{R})^{1/2} (\frac{\mu_{0} -
  \mu}{\sin\theta})(1 + \frac{\mu}{\mu_{0}})^{1/2}, \\ 
  & u_{\phi} = & (\frac{G M_{\ast}}{R})^{1/2}\frac{\sin\theta_{0}}{\sin\theta}
  (1 - \frac{\mu}{\mu_{0}})^{1/2}.\\
\end{array}
\label{eqn:infallvels}
\end{equation}

\noindent The density for the cloud material is then found by
integrating the mass flux along streamlines,
\begin{equation}
  \rho = \frac{-\dot{M}_{\rm infall}}{4\pi R^{2}u_{\rm R}}[1 + (3\mu_{0}^{2} -1) \zeta]^{-1}.
\label{eqn:density}
\end{equation}

The infalling cloud material is assumed to be molecular gas at a
temperature of 100 K.

\subsection{The winds}
\label{subsec:winds}

\begin{figure}
\begin{center}
    \begin{tabular}{c}
      \resizebox{80mm}{!}{{\includegraphics{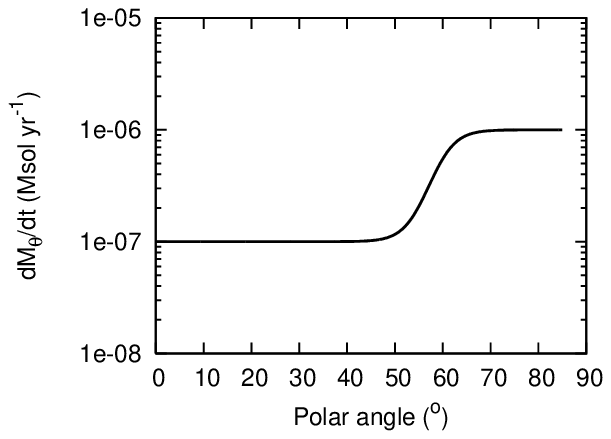}}} \\
      \resizebox{80mm}{!}{{\includegraphics{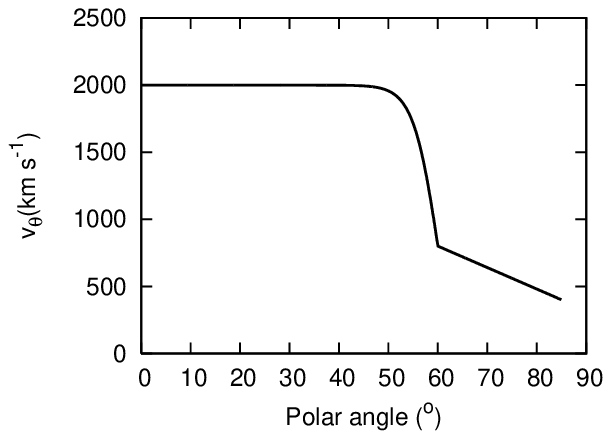}}} \\
    \end{tabular}
    \caption[]{Variation of the angle dependent mass-loss rate,
    $\dot{M}_{\theta}$ (upper panel) and wind velocity, $v_{\theta}$
    (lower panel) as a function of polar angle, $\theta$, for model
    R1.}
    \label{fig:mdotandv}
\end{center}
\end{figure}

Following the detection of equatorial winds from MYSOs
\citep{Drew:1993,Hoare:1994,Bunn:1995,Hoare:2006} we choose to
incorporate a disk wind into our model. The wind geometry we adopt is
aimed at representing the characteristics of disk wind models. In
particular we base our wind geometry on the hydrodynamical model of
\cite{Drew:1998} and the angle dependent model constructed by
\cite{Sim:2005}. In these works the disk wind is assumed to be
launched from the inner region ($< 10\;R_{\ast}$) of the accretion
disk. A considerable difficulty in constructing this model is the
range of scales across which important physics occurs. The cavity has
dimensions of the order of $10^{16}\cm$, whereas the acceleration
region of the winds is $\ltsimm 10^{12}\cm$. Therefore, to represent
the winds we have adopted a simple latitude dependent prescription,
where the winds are mapped on with terminal velocities, which aims to
capture the important features of a stellar/disk wind combination. The
mass-loss rate is then,

\begin{eqnarray}
   \dot{M}_{\theta} & = & \frac{\dot{M}_{\rm D}}{2} \left[1+\lambda + (1-\lambda)\tanh((\theta - \theta_{\rm S})\chi)\right], \\
   \lambda & = & \dot{M}_{\rm S}/ \dot{M}_{\rm D} \nonumber
\end{eqnarray}

\noindent $\dot{M}_{\theta}$ is the mass-loss rate of the outflow
(stellar and disk winds) as a function of polar angle, $\theta_{\rm
S}$ is the transition angle between the stellar and disk winds, and
$\chi$ is a steepening factor to sharpen the transition from the
stellar wind to the disk wind (consequently, the frictional heating
between the two flows is linked to this parameter). $\dot{M}_{\rm S}$
and $\dot{M}_{\rm D}$ would be the mass-loss rates of the stellar and
disk wind, respectively, if the mass-loss was isotropic and
spherical. Therefore, the \textit{actual} net mass-loss rates of the
stellar and disk winds can be found by integrating $\dot{M}_{\theta}$
over the appropriate angular range. We set $\chi = 5$. The wind
velocity as a function of polar angle, $v_{\theta}$, is,

\begin{equation}
v_{\theta} = \left\{
\begin{array}{lll}
\frac{v_{\rm D}}{3} \left[1 + \gamma + (1 - 2\gamma)\tanh((\theta - \theta_{\rm S})\chi)\right] & ;0  < \theta \ltsimm \theta_{\rm S} \\ 
 \frac{v_{\rm D}}{3}\left[1 + \gamma + (2 - \gamma)(\frac{\theta - \theta_{\rm S}}{\theta_{\rm D} - \theta_{\rm S}})\right]  & ;\theta_{\rm S} < \theta \ltsimm \theta_{\rm D}\\ 
0 & ;\theta_{\rm D}  < \theta \ltsimm \pi/2 \\

\end{array}\right.
\label{eqn:vels}
\end{equation}

\noindent where $\gamma = v_{\rm S}/v_{\rm D}$, $v_{\rm S}$ and
$v_{\rm D}$ are the terminal velocities of the stellar and disk winds
respectively, and $\theta_{\rm D}$ is the transition angle above which
there is no wind blown from the disk. The wind density as a function
of polar angle is then calculated from,
\begin{eqnarray}
\rho_{\theta} = \frac{\dot{M}_{\theta}}{4\pi R^{2}v_{\theta}} 
\end{eqnarray}

The angle dependent mass-loss rate and velocity profiles for model R1
are shown in Fig.~\ref{fig:mdotandv}. The winds are mapped into cells
within a small radius of the grid origin ($\simeq 1\times10^{15}\cm$)
provided that prior to the mapping a cell contains wind material
(i.e. not inflowing cloud material). When necessary, additional cells
in the first row above the disk plane (the $r$-axis in the model) are
used to map on the winds (i.e. when $\theta_{\rm D}\simeq
89^{\circ}$). This is to ensure that the angular dependence of the
disk wind is resolved sufficiently. The stellar and disk winds are
switched on at $t=0$. The unshocked winds are assumed to be ionized
gas at a temperature of $10^{4}\;$K.

\subsection{The cavity}

In the model the initially un-shocked stellar/disk winds are separated
from the infalling cloud material by a cavity wall. The shape and
position of the cavity wall is described by a simple analytical
prescription relating $z$ and $r$ on its surface \citep{alvarez:2004a},

\begin{eqnarray}
 & z = & \left[\left(\frac{r}{r_{\rm cav}}\right)^{\beta} - 1\right]\kappa,  \\
 {\rm where} & \kappa = & \frac{a \cos\omega}{(\frac{a}{r_{\rm cav}} 
   \sin\omega)^{\beta} - 1}.
\label{eqn:cavity}
\end{eqnarray}

\noindent $\omega$ is the half opening angle of the cavity, $\beta$
controls the initial curvature of the cavity wall ($\beta = 1$ or 2
corresponds to a cavity wall with either a straight or parabolic
shape), and $a$ is a characteristic scale size of the cavity (we
assume that $a=8\times10^{16}\;{\rm cm}$, approximately in agreement
with length scales for bi-polar outflows around MYSOs). It should be
noted that Eq.~\ref{eqn:cavity} is only suitable for $\omega \gtsimm
3^{\circ}$.

\subsection{The hydrodynamical code}

To perform the simulations we use the hydrodynamical code
\textsc{VH-1} \citep{Blondin:1990}. The code utilizes the Piecewise
Parabolic Method \citep{Colella:1984} to solve the gas dynamics
equations at cell boundaries. The version of the code we use in this
work has undergone some modifications which include: radiative cooling
\citep{Strickland:1995}, a direct Eulerian solver, and portability to
multi-processor machines using the message passing interface (MPI) via
the grid parallelization scheme described in \cite{Saxton:2005}. We
include radiative cooling for optically thin gas above $T=10^{4}\;$K,
and assume that molecular gas below this temperature instantaneously
cools to a temperature of $T=10^{2}\;$K. Gravity and rotation due to a
central point source are incorporated through effective potential
terms. The code also includes advected scalar fluid variables (dyes)
which are used to trace the position of the separate components of the
flow, i.e. stellar wind, disk wind, or cloud material.

We use a 2D, cylindrically symmetric, fixed grid with constant
resolution throughout the simulations of $\delta z = \delta r =
8. \dot{3}\times10^{13}\cm$, which for the grid of the fiducial model
(R1 in Table~\ref{tab:models}) equates to $r\times z =
(5\times10^{16}\cm)\times(8\times10^{16}\cm) = 600\times960\;{\rm
cells}$. The boundary conditions (BCs) are set as follows: the $r=0$
boundary uses a symmetric BC, the $z=0$ boundary uses an outflow BC
(to allow gas to flow onto a hypothetical disk). The $z=z_{\rm max}$
and $r=r_{\rm max}$ boundaries use either a zero-gradient outflow BC
if the cell adjacent to the boundary is wind material or a fixed
inflow condition for cloud material if the adjacent cell contains
cloud material. In the latter case Eqs.~\ref{eqn:infallvels} and
\ref{eqn:density} are used to populate the ghost cells.

\vspace{-3mm}
\subsection{X-ray emission}
\label{subsec:xrayemission}

To allow a comparison to be made between \textit{Chandra} X-ray
observations of MYSOs and our simulations we calculate attenuated
X-ray fluxes. Radiative transfer calculations are performed on the 2D
models using the method described in \cite{Dougherty:2003}. This
involves calculating emission and absorption coefficients for each
cell on the grid and integrating the radiative transfer equation. For
the emissivities we use values for optically thin gas in collisional
ionization equilibrium obtained from look-up tables calculated from
the \textsc{MEKAL} plasma code \citep[][and references
there-in]{Liedahl:1995} containing 200 logarithmically spaced energy
bins in the range 0.1-10 keV, and 101 logarithmically spaced bins in
the range $10^{4}-10^{9}\;$K. Solar abundances are assumed. The
advected fluid variables (dyes) are used to identify the X-ray
emission contributions from the stellar and disk winds and the heated
cloud material.

\begin{figure*}
\begin{center}
    \begin{tabular}{ccc}
      \resizebox{40mm}{!}{{\includegraphics{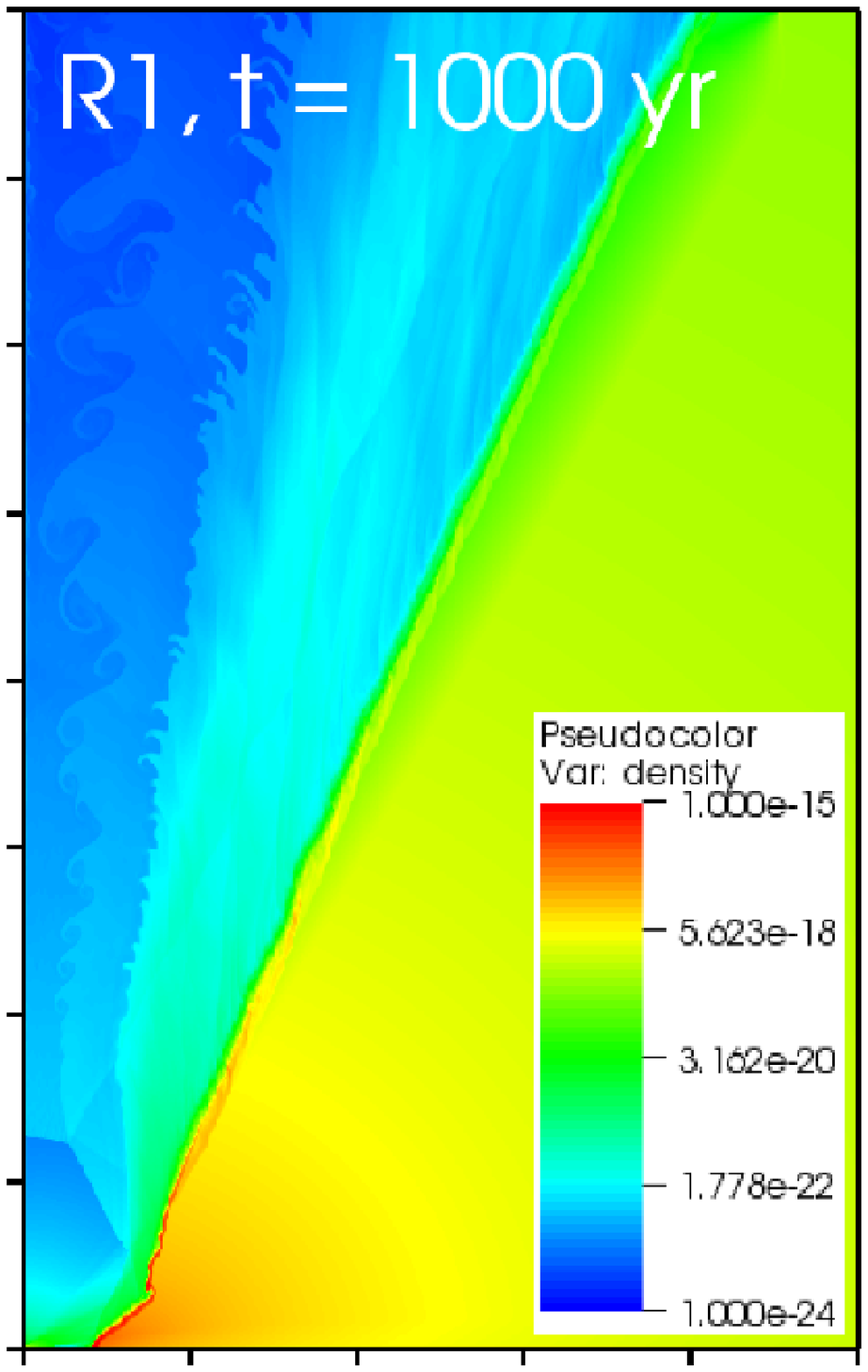}}} &
      \resizebox{40mm}{!}{{\includegraphics{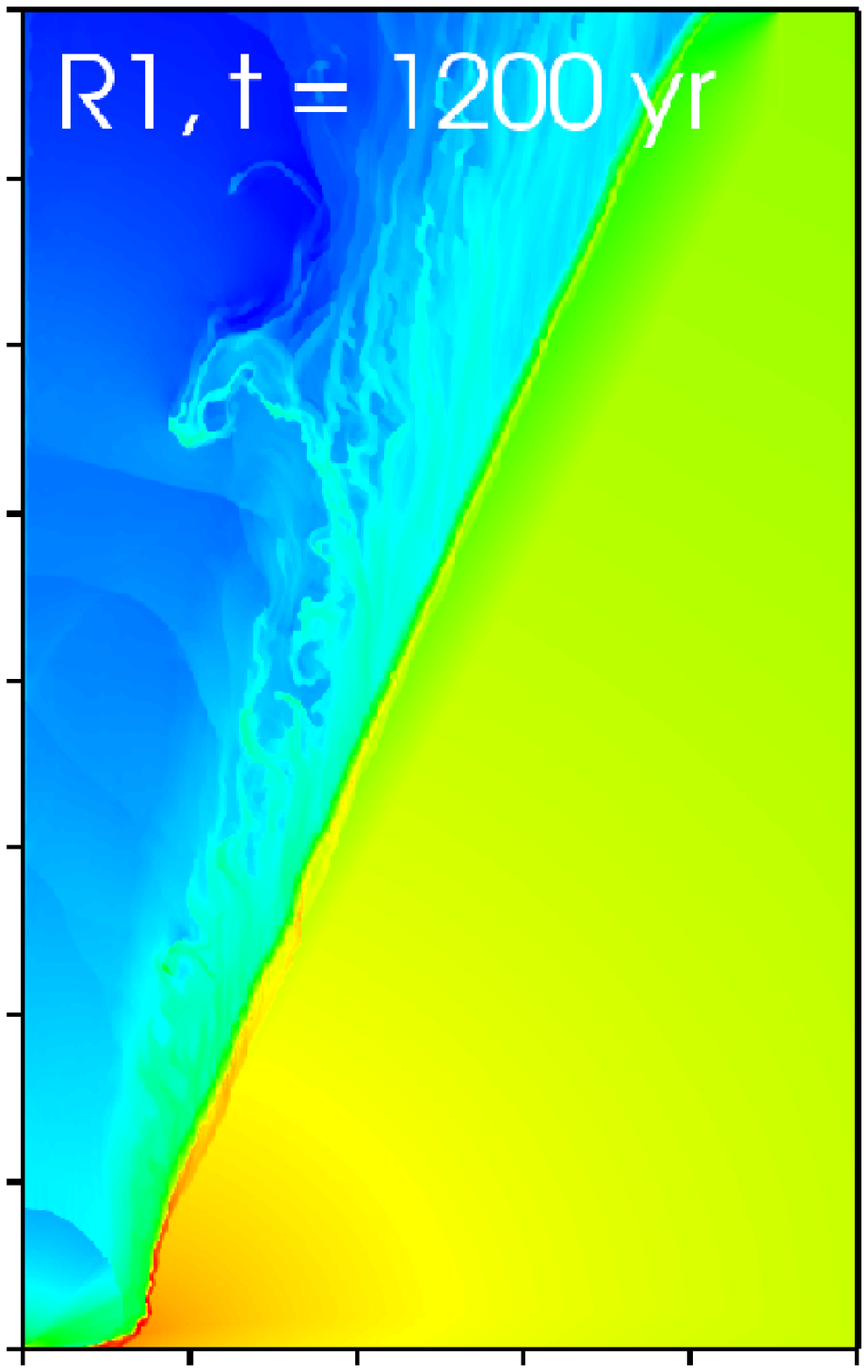}}} &
      \resizebox{40mm}{!}{{\includegraphics{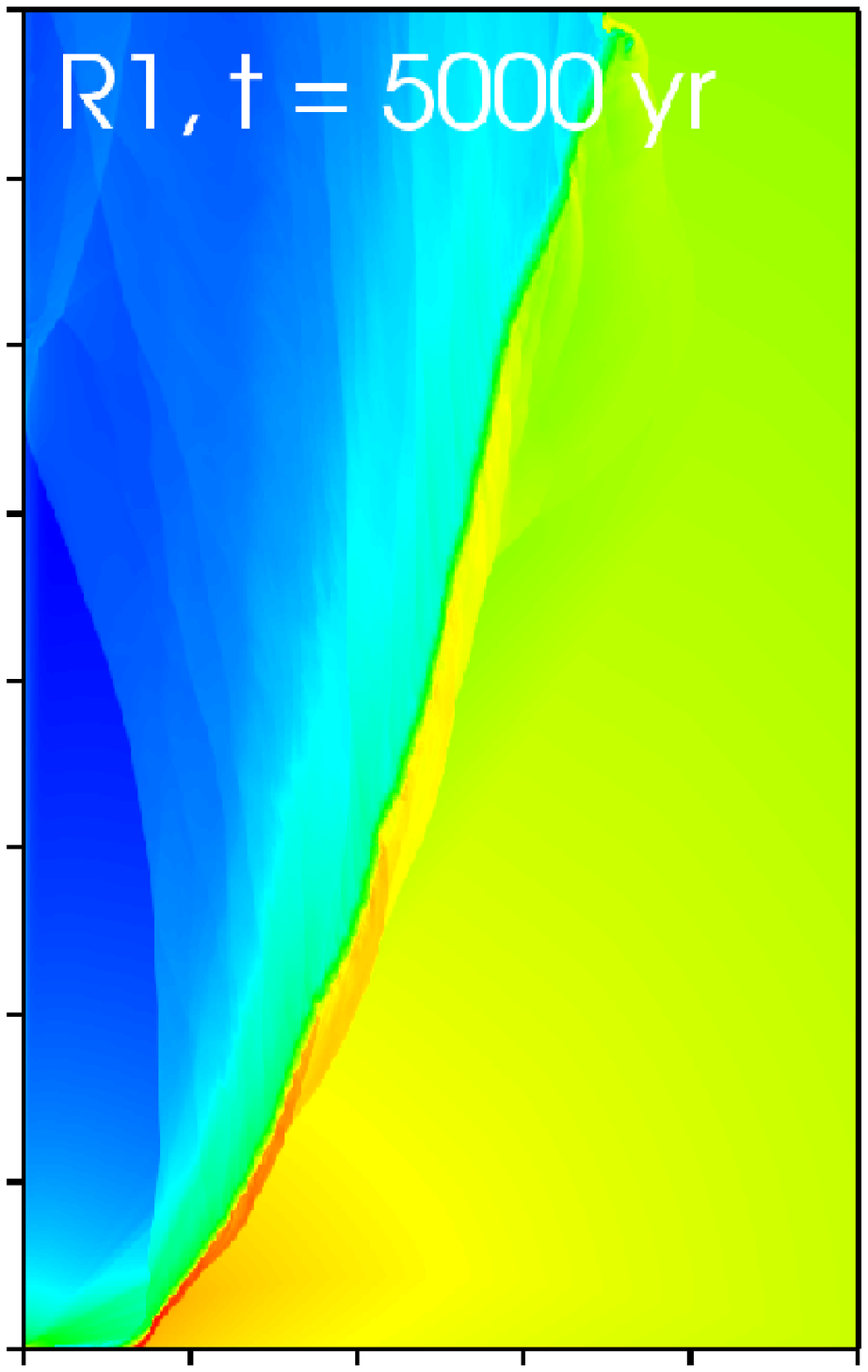}}} \\

      \resizebox{40mm}{!}{{\includegraphics{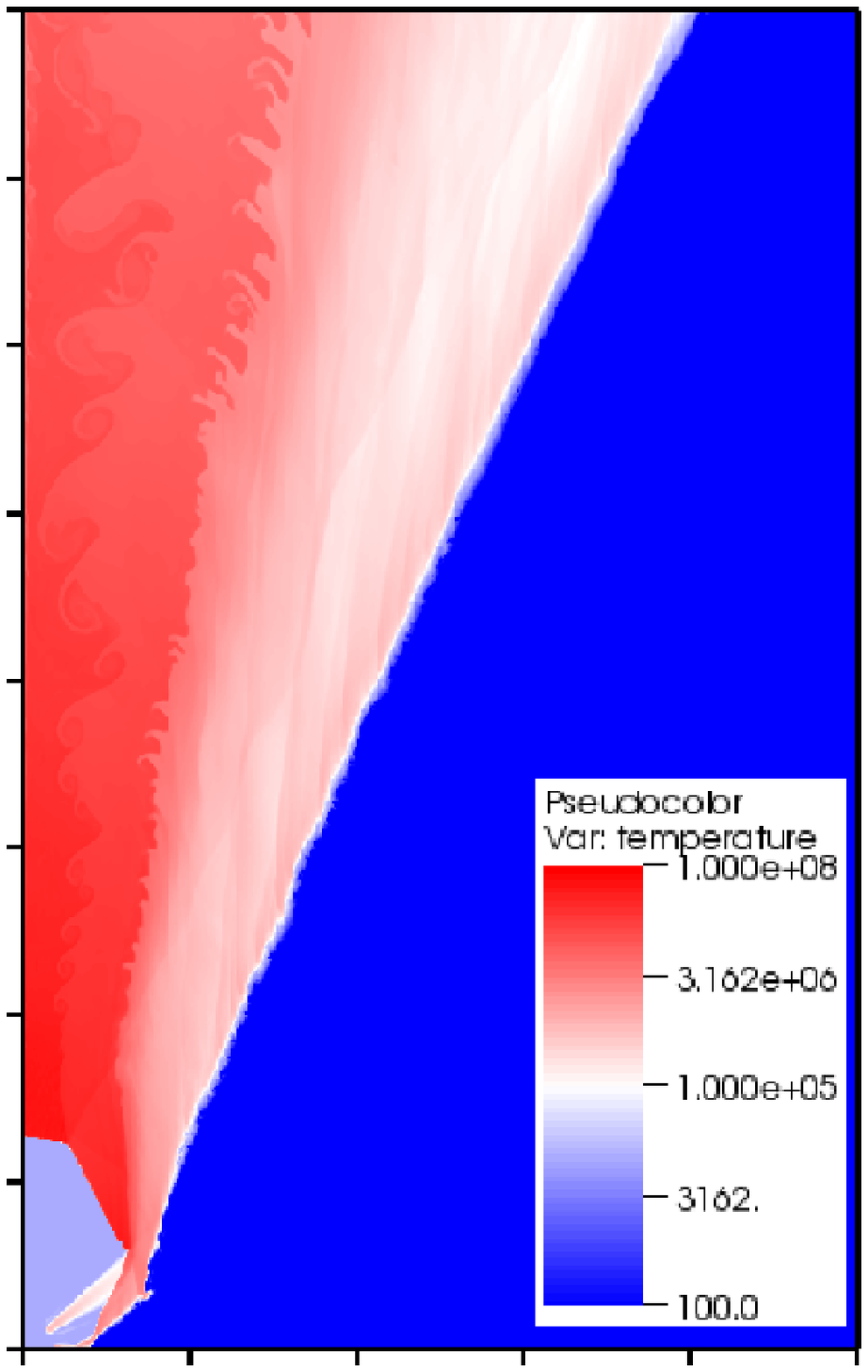}}} &
      \resizebox{40mm}{!}{{\includegraphics{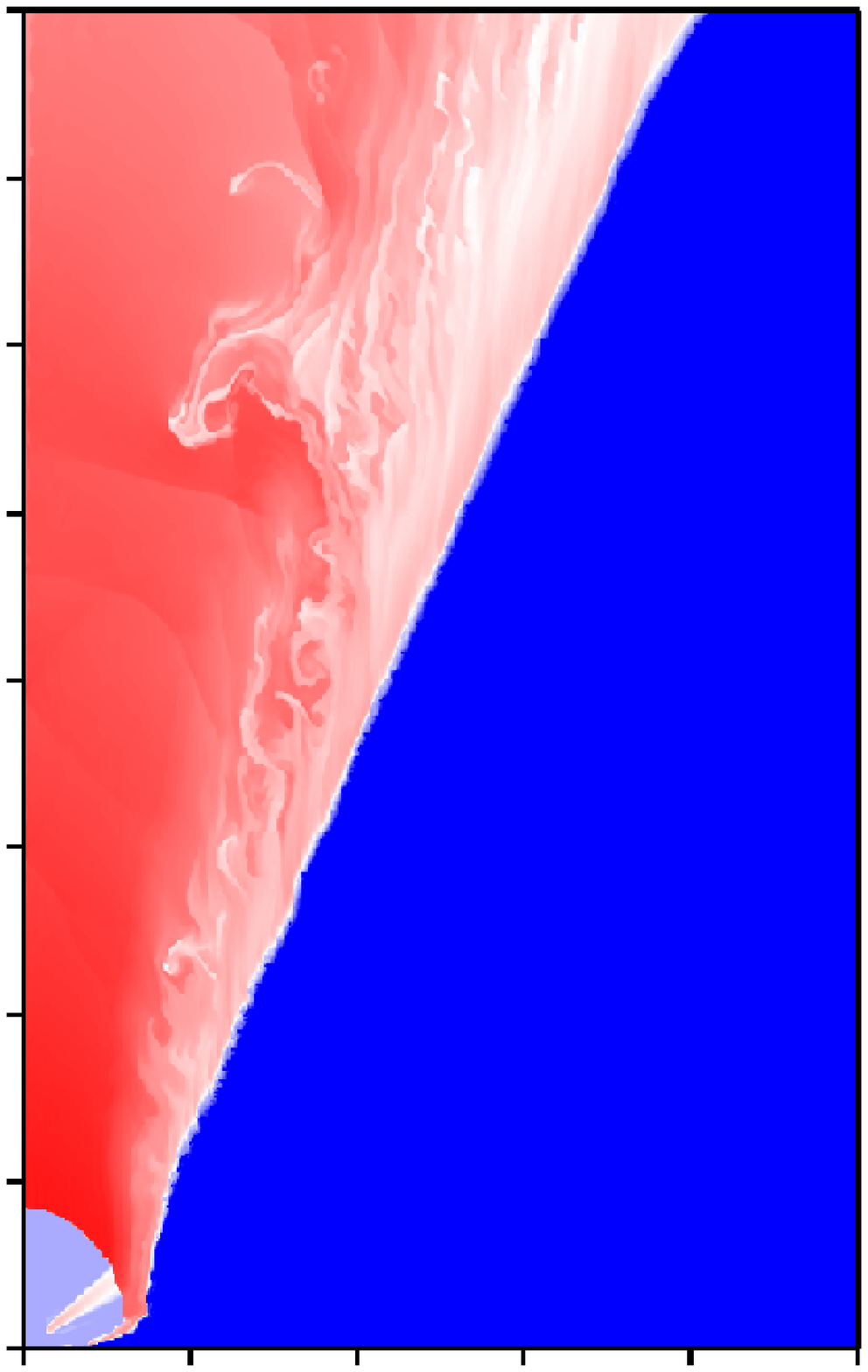}}} &
      \resizebox{40mm}{!}{{\includegraphics{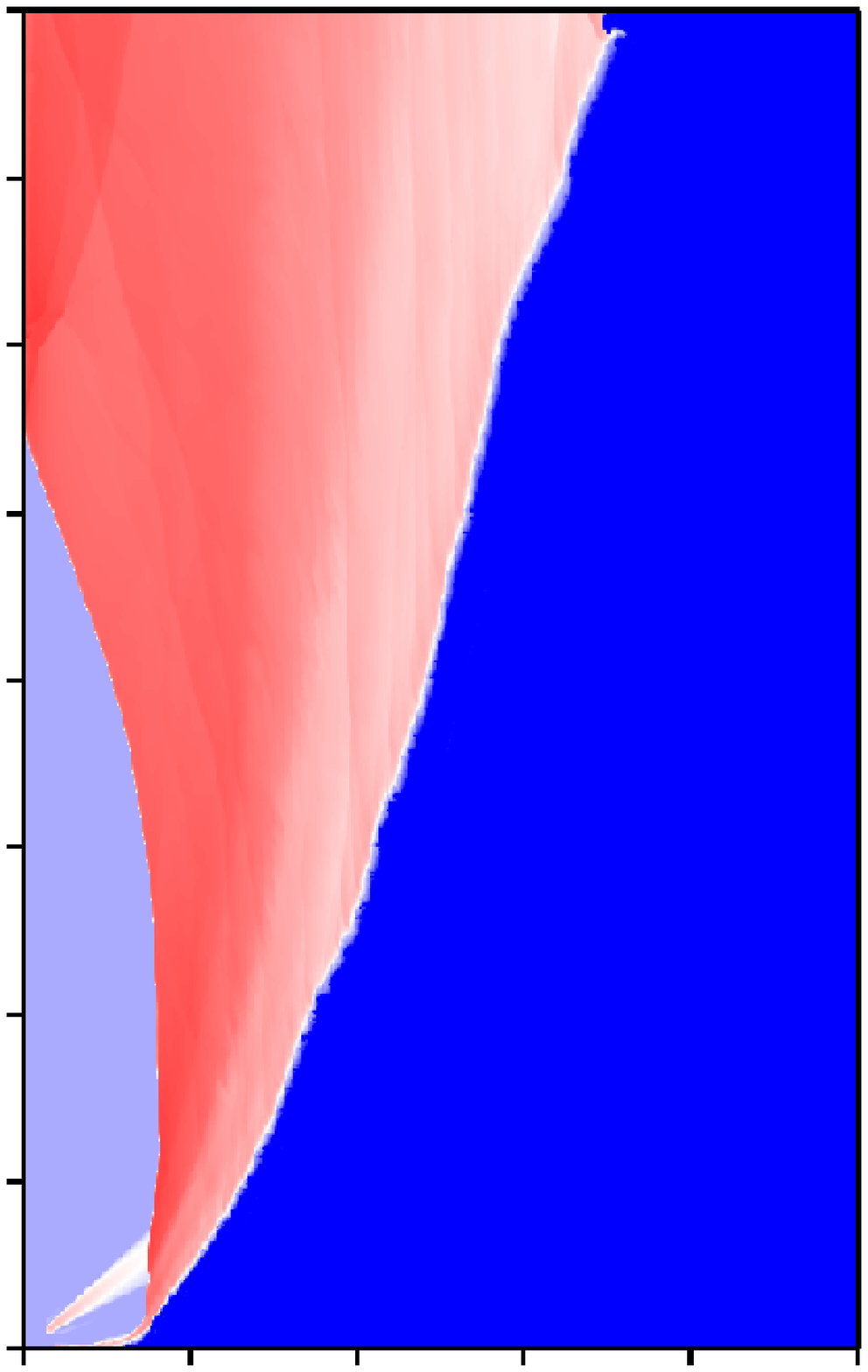}}} \\

    \end{tabular}
    \caption[]{Snapshots of density (upper panels) and temperature
    (lower panels) taken from model R1 at (from left to right):
    $t=1000$, 1200, and 5000 yr. The tick marks on the axis correspond
    to a distance of $10^{16}\;$cm. The strip of $T\sim10^{5}\;$K gas
    which extends diagonally from the wind source is due to frictional
    heating at the interface between the fast stellar wind and the
    slower disk wind.}
    \label{fig:cavevo}
\end{center}
\end{figure*}

For lines-of-sight which exit the grid while in infalling cloud
material we calculate an additional column density beyond the grid
boundary by numerically integrating Eq.~\ref{eqn:density}. X-rays
which pass through the disk plane within the cavity radius, $r_{\rm
cav}$, are assumed to be occulted by the accretion disk. For certain
viewing angles there are lines-of-sight which exit the grid while in
the stellar or disk winds. The winds have a much lower density than
the cloud material, and therefore contribute far less absorption. The
model cannot be used to calculate the attenuated emission for
inclination angles smaller than the cavity opening angle due to
insufficient extent of the computational grid (in reality there is
likely to be additional absorption from swept-up cloud material
formerly in the cavity).

Once model fluxes have been calculated we use the \textit{Chandra}
ACIS-I effective
area\footnote{http://cxc.harvard.edu/cgi-bin/build\_viewer.cgi?ea} to
convert the synthetic spectra into counts space. When calculating the
synthetic fluxes we assume a distance to the MYSO of 1 kpc and an ISM
column (between the observer at Earth and the edge of the cloud) of
$1.9\times10^{21}\cm^{-2}$.

\begin{table*}
\begin{center}
  \caption[]{List of the parameters used in the simulations. $\beta$
    is the parameter which defines the curvature of the cavity wall
    (Eq.~\ref{eqn:cavity}), $r_{\rm c}$ is the centrifugal radius,
    $\omega$ is the cavity half opening angle, $\dot{M}_{\rm infall}$
    is the mass infall rate of the molecular cloud, $\theta_{\rm S}$
    is the transition angle between the stellar and disk wind, and
    $\theta_{\rm D}$ is the limiting angle above which there is no
    wind (i.e. only an accretion flow). For the wind terminal
    velocity, $v$, mass-loss rate, $\dot{M}$, and power, $\zeta$, the
    subscripts ``S'' and ``D'' refer to the stellar and disk wind
    components, respectively. $M_{\ast} = 10\thinspace \Msol$ and
    $r_{\rm cav}=3.0\times10^{15}\cm$ in all models, except models
    AFGL 2591 and Mon R2 where $r_{\rm cav}= 4.5\times10^{15}$ and
    $1.5\times10^{15}\cm$, respectively.}
\begin{tabular}{lclcccccccccc}
\hline
Model & $\beta$ & $r_{{\rm c}}$ & $\omega$ & $\dot{M}_{\rm infall}$ & $\theta_{\rm S}$ & $\theta_{\rm D}$ & $v_{\rm S}$ & $v_{\rm D}$ & $\dot{M}_{\rm S}$ & $\dot{M}_{\rm D}$ & $\zeta_{\rm S}$ & $\zeta_{\rm D}$\\
 &  & (cm) & $(^{\circ})$ & ($\Msolpyr$) & $(^{\circ})$ & $(^{\circ})$ & (km s$^{-1}$) & (km s$^{-1}$) & ($\Msolpyr$) & ($\Msolpyr$) & (L$_{\odot}$) & (L$_{\odot}$) \\
\hline
R1 & 1 & $3\times10^{15}$ & 30 & $2\times10^{-4}$ & 60 & 85 & 2000 & 400 & $1\times10^{-7}$ & $1\times10^{-6}$ & 250 & 132 \\
\hline
R2 & 1 & $3\times10^{15}$ & 30 & $1\times10^{-4}$ & 60 & 85 & 2000 & 400 & $1\times10^{-7}$ & $1\times10^{-6}$ & 250 & 132  \\
R3 & 1 & $3\times10^{15}$ & 30 & $1\times10^{-3}$ & 60 & 85 & 2000 & 400 & $1\times10^{-7}$ & $1\times10^{-6}$ & 250 & 132 \\
R4 & 1 & $3\times10^{15}$ & 30 & $2\times10^{-4}$ & 60 & 85 & 2000 & 400 & $1\times10^{-6}$ & $1\times10^{-6}$ & 1674 & 158 \\
R5 & 1 & $3\times10^{15}$ & 30 & $2\times10^{-5}$ & 60 & 85 & 2000 & 400 & $1\times10^{-7}$ & $1\times10^{-7}$ & 167 & 16 \\
R6 & 1 & $3\times10^{15}$ & 30 & $1\times10^{-3}$ & 60 & 85 & 2000 & 400 & $1\times10^{-7}$ & $2\times10^{-6}$ & 343 & 261 \\
R7 & 1 & $3\times10^{15}$ & 30 & $2\times10^{-4}$ & 60 & 85 & 2000 & 400 & $5\times10^{-8}$ & $1\times10^{-6}$ & 171 & 131 \\
\hline
R8 & 1 & $3\times10^{15}$ & 30 & $2\times10^{-4}$ & 60 & 85 & 3000 & 400 & $1\times10^{-7}$ & $1\times10^{-6}$ & 555 & 218\\
R9 & 1 & $3\times10^{15}$ & 30 & $2\times10^{-4}$ & 60 & 85 & 1000 & 400 & $1\times10^{-6}$ & $1\times10^{-6}$ & 427 & 80 \\
R10 & 1 & $3\times10^{15}$ & 30 & $2.5\times10^{-5}$ & 60 & 85 & 1000 & 400 & $2\times10^{-8}$ & $2\times10^{-7}$ & 12 & 15 \\
\hline
R11 & 1 & $3\times10^{15}$ & 30 & $2\times10^{-4}$ & 70 & 85 & 2000 & 400 & $1\times10^{-7}$ & $1\times10^{-6}$ & 322 & 75 \\
R12 & 1 & $3\times10^{15}$ & 30 & $2\times10^{-4}$ & 85 & 89 & 2000 & 400 & $1\times10^{-7}$ & $1\times10^{-6}$ & 433 & 21 \\
R13 & 1 & $3\times10^{15}$ & 30 & $2\times10^{-4}$ & $-$ & $-$ & 2000 & $-$ & $1\times10^{-7}$ & $-$ & 470 & $-$ \\
\hline
R14 & 1 & $3\times10^{15}$ & 5 & $2\times10^{-4}$ & 60 & 85 & 2000 & 400 & $1\times10^{-7}$ & $1\times10^{-6}$ & 250 & 132\\
R15 & 1 & $3\times10^{15}$ & 10 & $2\times10^{-4}$ & 60 & 85 & 2000 & 400 & $1\times10^{-7}$ & $1\times10^{-6}$ & 250 & 132\\
R16 & 1 & $3\times10^{15}$ & 45 & $2\times10^{-4}$ & 60 & 85 & 2000 & 400 & $1\times10^{-7}$ & $1\times10^{-6}$ & 250 & 132\\
R17 & 1 & $3\times10^{15}$ & 60 & $2\times10^{-4}$ & 60 & 85 & 2000 & 400 & $1\times10^{-7}$ & $1\times10^{-6}$ & 250 & 132\\
\hline
R18 & 2 & $3\times10^{15}$ & 30 & $2\times10^{-4}$ & 60 & 85 & 2000 & 400 & $1\times10^{-7}$ & $1\times10^{-6}$ & 250 & 132\\
\hline
R19 & 1 & $8\times10^{14}$ & 30 & $2\times10^{-4}$ & 60 & 85 & 2000 & 400 & $1\times10^{-7}$ & $1\times10^{-6}$ & 250 & 132\\
R20 & 1 & $6\times10^{15}$ & 30 & $2\times10^{-4}$ & 60 & 85 & 2000 & 400 & $1\times10^{-7}$ & $1\times10^{-6}$ & 250 & 132\\
\hline
S106 & 1 & $3\times10^{15}$ & 50 & $3\times10^{-5}$ & 60 & 85 & 1000 & 350 & $4\times10^{-8}$ & $4\times10^{-7}$ & 24 & 25 \\   
Mon R2 & 1 & $8\times10^{14}$ & 10 & $2\times10^{-5}$ & 60 & 85 & 2000 & 400 & $1\times10^{-8}$ & $1\times10^{-7}$ & 23 & 14 \\   
AFGL 2591 & 1 & $4.5\times10^{15}$ & 15 & $3\times10^{-4}$ & 60 & 85 & 2000 & 400 & $1\times10^{-7}$ & $1\times10^{-6}$ &  250 & 132 \\ 
\hline
\label{tab:models}
\end{tabular}
\end{center}
\end{table*}

\vspace{-2mm}
\section{Results}
\label{sec:results}

To explore the impact of varying the different parameters in the model
on both the evolution of the cavity and the resulting X-ray emission
we have performed an extensive parameter space exploration
(Table~\ref{tab:models}). Each simulation is allowed to run until a
model time of 2000 yr, except model R1 which was allowed to run until
a model time of 5000 yr. We first describe the cavity evolution and
X-ray emission features of the fiducial model (R1), and then examine
the influence of the various parameters on the results.

\subsection{The standard model}

\subsubsection{Dynamical and X-ray properties}

The standard model (R1) mimics the MYSO disk wind model of
\cite{Drew:1998}, with identical wind geometry and stellar and disk
wind velocities. The disk wind mass-loss rate is chosen to match
observations \citep{Felli:1984}. The stellar wind speed and mass-loss
rate are representative of an O8-9V star\footnote{The stellar mass
adopted in the simulations of $10\Msol$ is lower than that expected
for an 08-9V star ($\sim 30\Msol$). An increase in the mass of the
central star will increase the gravitational attraction felt by the
infalling cloud material, and thus the infall velocity will increase
and the density will be decreased. However, tests show that this
difference does not significantly affect the results.}
\citep{Howarth:1989}. The mass infall rate for accretion onto the
disk, $\dot{M}_{\rm infall}$, is chosen to be intermediate between the
value of $1.1\times10^{-4}\Msolpyr$ derived by \cite{alvarez:2004a}
for the MYSO Mon R2 IRS 3A, and theoretically predicted mass infall
rates of $\simeq 10^{-3}\Msolpyr$ for massive stars
\citep{Banerjee:2007,Krumholz:2009}. The centrifugal radius of the
cloud, $r_{\rm c}$, is taken to be of the order of the observed disk
radius for MYSOs \citep[e.g. ][]{alvarez:2004a}. As a simple first
approximation the cavity radius, $r_{\rm cav}$, is taken to be equal
to the centrifugal radius of the cloud, and the cavity is assumed to
be conical ($\beta=1$).

Fig.~\ref{fig:cavevo} shows the spatial distribution of material in
the simulation; the disk wind lines the cavity wall and separates the
stellar wind from the infalling molecular cloud material. Close to the
star there is a small strip of frictionally heated wind gas which
reaches temperatures of $\sim 10^{5}\;$K and therefore emits soft
X-rays ($E\sim 0.1\;$keV), although with a negligible observed flux
when compared to the shocked winds. The stellar and disk winds in the
simulations are supersonic, so that flow incident against the cavity
wall results in shocked gas bounded by a reverse shock (this is best
seen in the temperature images in Fig.~\ref{fig:cavevo}). The
postshock temperature is dictated by the preshock wind speed normal to
the shock. For $v_{\rm S}=2000\;\kmps$, temperatures up to
$T\sim10^{8}\;$K are obtained, whereas the slower disk wind ($v_{\rm
D}=400\;\kmps$) is shocked only up to temperatures
$\sim10^{6.5}\;$K. At the base of the cavity wall the small angle of
incidence to the normal for the winds, and thus small angle of
reflection, causes the shocked winds to intersect the rotational
symmetry axis and close off the unshocked stellar wind. A pressure
balance between the shocked and unshocked gas causes the reverse shock
to continually enclose the latter. The reverse shock is almost normal
to the preshock flow near the disk plane and on the symmetry axis
where the highest postshock temperatures are obtained. Because the ram
pressure of the inflow/outflow is angle dependent and the base of the
cavity is subject to instabilities the shape of the reverse shock is
often non-spherical. In Fig.~\ref{fig:cavevo} the reverse shock can be
seen in the bottom left of the grid, where the enclosed preshock wind
has temperature $T=10^{4}\;$K. The position of the reverse shock
oscillates and its shape changes with time, due to small fluctuations
in the shape and size of the base of the cavity wall as inflowing
material is ablated and incorporated into the outflow, and as new
inflowing material replenishes it. The size of the reverse shock is
tied to the postshock pressure. The oscillations have a large effect
on the observed emission as the postshock density, and therefore the
emission measure, are dependent on the shock position. The postshock
stellar wind has velocity vectors which are preferentially aligned in
the polar direction. Small kinks to the shape of the reverse shock
cause shear velocities in the post-shock flow, and KH instabilities
are produced which can clearly be seen at $t=1000\;$yrs. The shear
layer between the stellar and disk winds provides a site for the
growth of $\sim 10^{16}\;$cm amplitude KH instabilities on timescales
of $\sim$ a few years. By $t=1200\;$yrs an instability of this
proportion can be seen driving a clump of disk wind material into the
path of the stellar wind, which leads to mass-loading of the latter
\citep[for a review of mass-loading processes
see][]{Pittard:2007book}.

On larger scales the half opening angle of the cavity, $\omega$,
decreases until a temporary pressure balance is attained between the
gas on either side of the cavity wall at $t\simeq1000\;$yr. A dense
layer of compressed cloud material which grows with time results from
the pile-up of inflowing cloud material. At the top of the cavity a
small narrowing has occured due to the pressure difference across the
cavity wall; if no winds are blown at all the cavity closes up over
the star after $\sim 500\;$yrs. Tests performed using the streamline
equation (Eq.~\ref{eqn:streamline}) to calculate the initial cavity
shape and position did not prevent this occurence. By a simulation
time of $t=2000\;$yr, the cavity has widened slightly at the base. The
inflowing cloud material can still reach the disk plane and it is
conceivable that accretion onto a disk could still be on-going while
the winds are interacting with the cavity.

\begin{figure}
\begin{center}
    \begin{tabular}{c}
      \resizebox{70mm}{!}{{\includegraphics{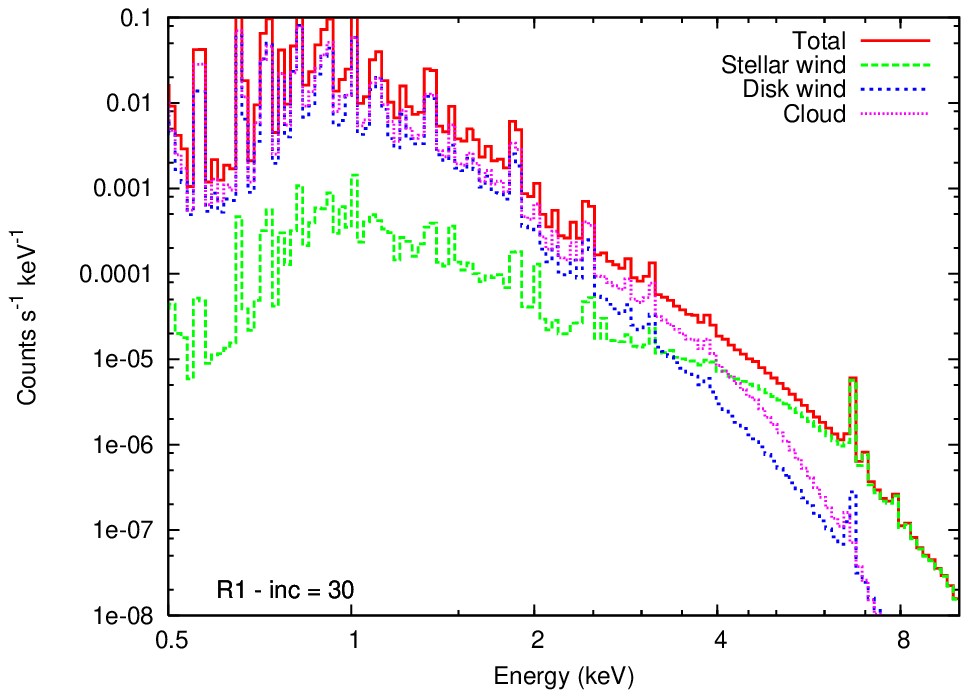}}} \\
      \resizebox{70mm}{!}{{\includegraphics{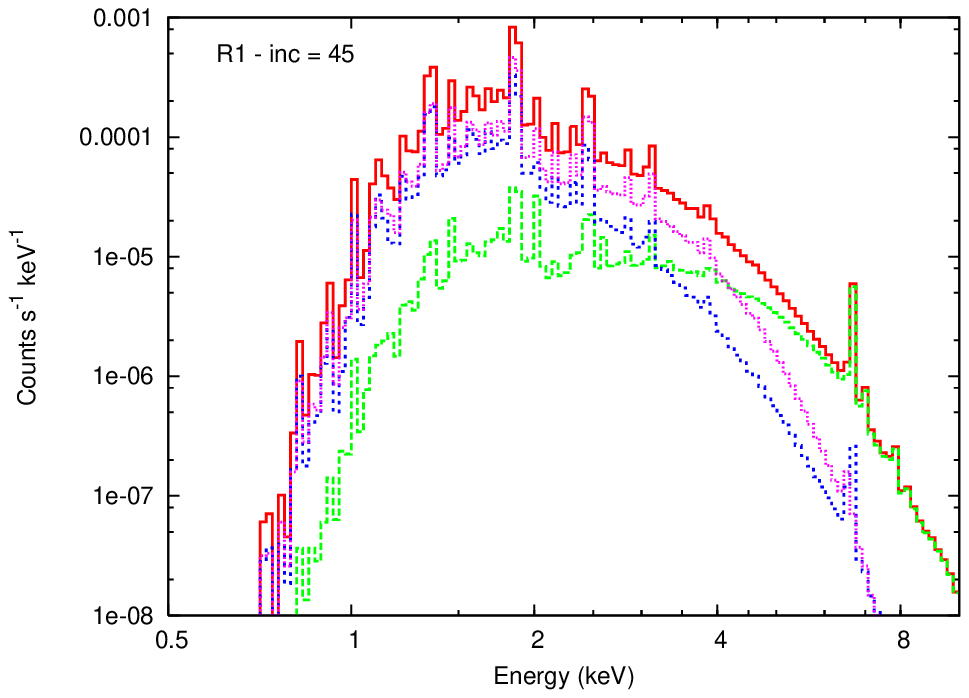}}} \\
      \resizebox{70mm}{!}{{\includegraphics{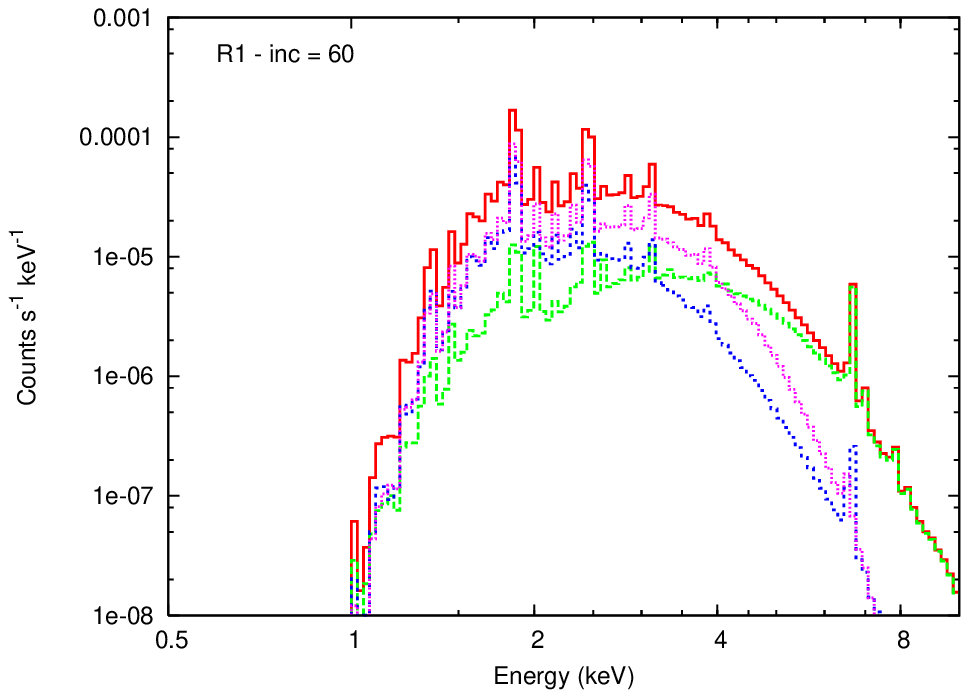}}} \\
      \resizebox{70mm}{!}{{\includegraphics{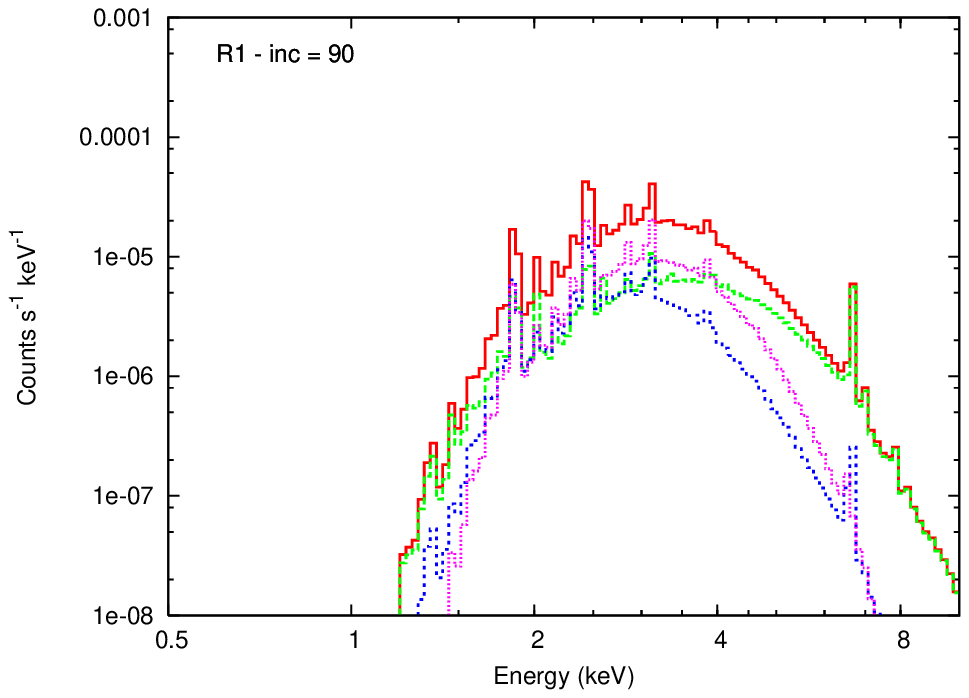}}} \\
    \end{tabular}
    \caption[]{Attenuated 0.5-10 keV X-ray spectra for model R1 at
    various inclination angles. From top to bottom: $i = 30, 45, 60$,
    and $90^{\circ}$. The total (red), stellar wind (green), disk wind
    (blue), and heated cloud material (pink) emission are shown. All
    spectra were calculated at $t = 1000\thinspace{\rm yr}$ and have
    been convolved with the \textit{Chandra} effective area. Note the
    difference in scale in the top panel.}
    \label{fig:incspectra}
\end{center}
\end{figure}

\begin{figure}
\begin{center}
    \begin{tabular}{cc}
      \resizebox{70mm}{!}{{\includegraphics{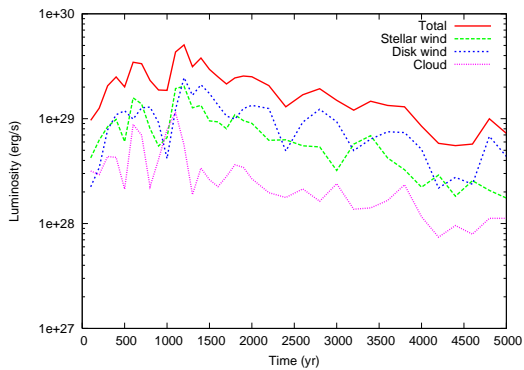}}} \\
    \end{tabular}
    \caption[]{Attenuated 0.1-10 keV lightcurve for model R1 with
    $i=60^{\circ}$. The lightcurves of the separate emission
    components (stellar wind, disk wind, cloud) are also plotted.}
    \label{fig:lcurves}
\end{center}
\end{figure}

\begin{figure}
\begin{center}
    \begin{tabular}{c}
      \resizebox{70mm}{!}{{\includegraphics{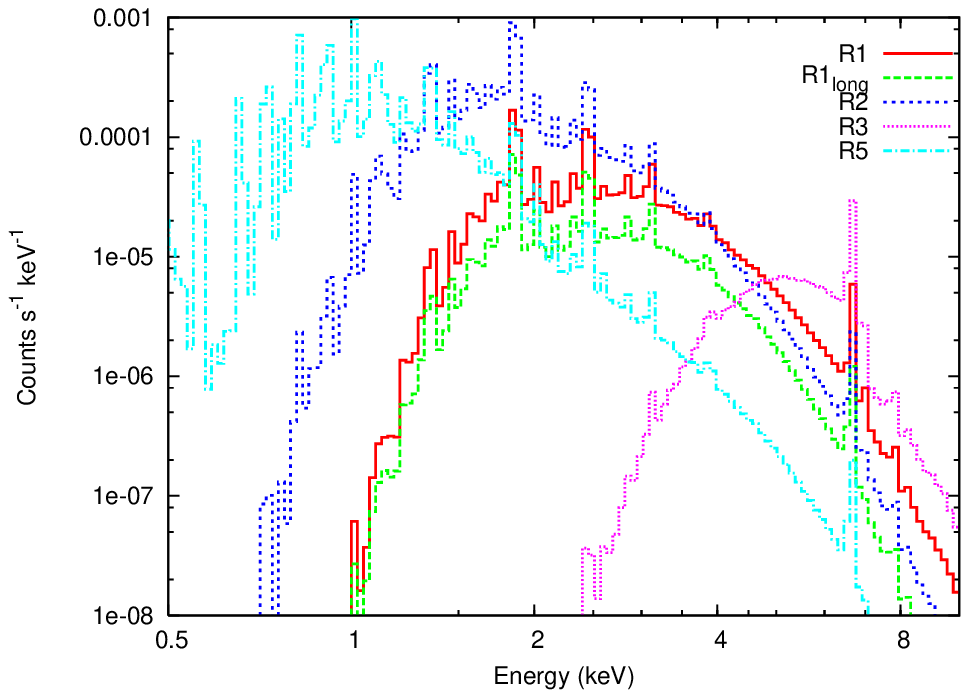}}} \\   
      \resizebox{70mm}{!}{{\includegraphics{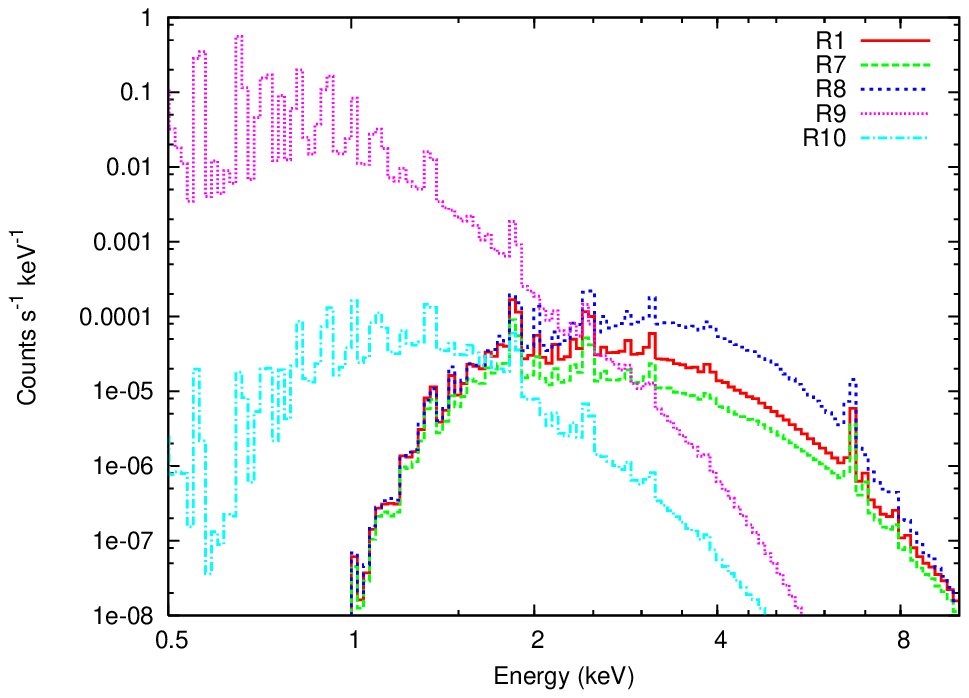}}} \\   
      \resizebox{70mm}{!}{{\includegraphics{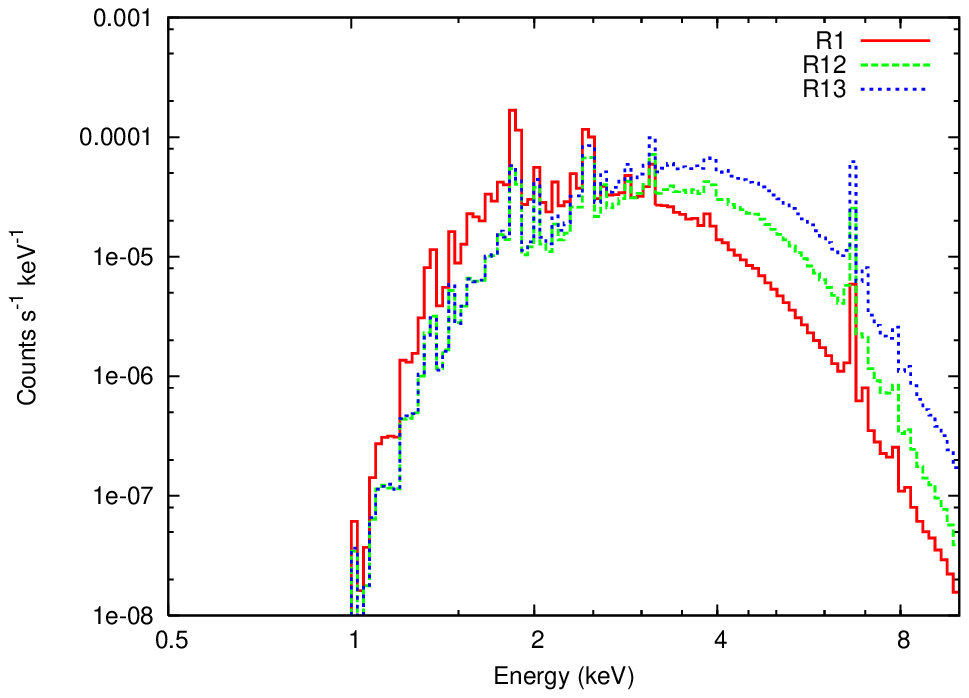}}} \\   
      \resizebox{70mm}{!}{{\includegraphics{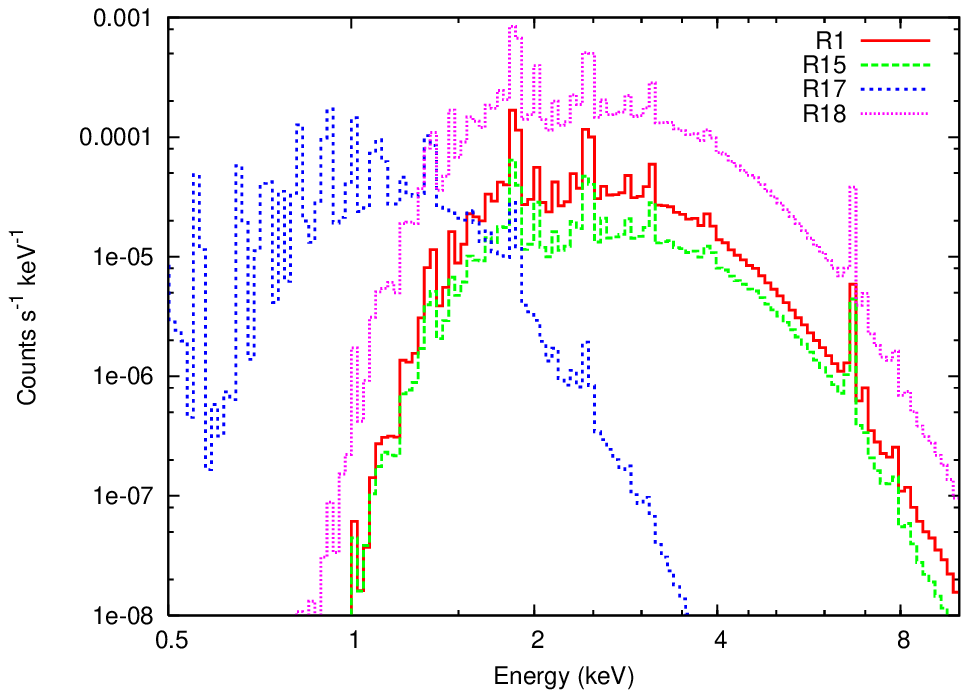}}} \\   
    \end{tabular}
    \caption[]{Total attenuated 0.5-10 keV spectra for
    simulations. The top panel shows R1, R1$_{\rm long}$, R2, R3, R5,
    the upper middle panel shows R1, R7, R8, R9, R10, the lower middle
    panel shows R1, R12, R13, and the lower panel shows R1, R15, R17,
    R18. All spectra were calculated at a time of $t =
    1000\thinspace{\rm yr}$ (except models R1$_{\rm long}$ which was
    at $t=5000\;$yr, and R3 which was at $t=$500 yrs), using an
    inclination angle, $i =60^{\circ}$, and have been convolved with
    the \textit{Chandra} effective area. Note the difference in scale
    in the upper middle panel.}
    \label{fig:spectra}
\end{center}
\end{figure}

Model R1 was allowed to run for an extended simulation time of 5000
yr. At $t>2000\;$yr the shape of the cavity wall continues to
change. The base of the cavity wall becomes more curved as time goes
on, and the cavity wall at the top of the grid narrows further
(Fig.~\ref{fig:cavevo}). The spectrum at $t=5000\;$yrs has a very
similar shape to that at $1000\;$yrs, though with a factor of $\sim 2$
lower normalization (Fig.~\ref{fig:spectra}). Similarly, the
\textit{Chandra} count rate is lower (model R1$_{\rm long}$ in
Table~\ref{tab:xrays}). These differences are unsurprising as the
reverse shock is now further from the star.

The X-ray emissivity of the shocked gas varies enormously with
position. The intrinsic X-ray emission comes mainly from disturbed
cloud material, then the disk wind, and finally the stellar wind
(Table~\ref{tab:xrays}). The shock driven into the cloud is too slow
to heat gas up to X-ray emitting temperatures, but large quantities of
cloud material are ablated by the outflowing disk wind and mixed into
this hotter flow (note that the cloud material does not directly
mass-load the stellar wind - Fig.~\ref{fig:cavevo} shows that it is
the disk wind which does this). This process heats the cloud material
to temperatures where (soft) X-rays are emitted. This process of
``mass-loading'' the outflowing disk wind creates densities which are
typically two orders of magnitude higher than the shocked stellar wind
at comparable heights above the disk plane. However, the difference
between the \textit{attenuated} luminosities of the cloud and stellar
wind components is very small ($L_{\rm att_{C}}=8\times10^{28}$ and
$L_{\rm att_{S}}=7\times10^{28}\ergps$, respectively). The explanation
is that the stellar wind emission is harder and extends to higher
energies, and is less affected by attenuation. In contrast, the cloud
material, which is heated to lower temperatures, emits prolifically at
low energies and has a much higher intrinsic luminosity, but its
emission is subject to considerable attenuation at $E \ltsimm
1\;$keV. Examining the attenuated spectrum at $i=60^{\circ}$
(Fig.~\ref{fig:incspectra}), we find that the stellar wind material
dominates the emission at $E\gtsimm 4\;$keV, with comparable
contributions from the winds and cloud material below this energy.

The observed model X-ray emission for $i=60^{\circ}$ shows variability
on timescales of order a few hundred years
(Fig.~\ref{fig:lcurves}). The total emission initially rises after the
stellar and disk winds are switched on and a reverse shock is allowed
to develop. There are then quasi-periodic oscillations, with
noticeable dips at $t\simeq500, 1000, 1300,$ and 1700 yrs. The
fluctuation in the total emission approximately follows the behaviour
of the emission from the stellar wind. Some features are due to the
disk wind luminosity, which is sometimes the dominant
emitter. Fluctuations in both of these emission components is strongly
related to the position of the reverse shock. At $t=500\;$yr the
distance of the reverse shock from the star increases, and a reduction
of the stellar wind luminosity occurs. Similarly, the maximum total
luminosity occurs at $t=1200\;$yr, in conjunction with a minimum in
the distance of the reverse shock from the star and a shape which
results in more stellar wind material being shocked to the highest
temperatures (Fig.~\ref{fig:cavevo}). The variability in the observed
emission is a general feature for all the simulations and complicates
comparisons between the different model results. In simulation R1 the
cavity has reached an approximate steady state and therefore the
emission averaged over 500 or 1000 yrs should remain roughly
constant. At $t=1200\;$yr a clump of disk wind is driven into the
stellar wind by turbulent mixing (Fig.~\ref{fig:cavevo}). Such mixing
boosts the observed emission of both winds by heating the cooler disk
wind and increasing the emission measure of the stellar wind. However,
the bulk of the emission still originates from close to the reverse
shock. The temperature of the stellar wind gas is also increased as it
impacts the clumps and additional shocks are produced. At $t>2000\;$yr
the attenuated luminosity appears to steadily decrease and the
amplitude of the fluctuations seem reduced. This suggests that the
variability at $t<2000\;$yr may be related to an initial evolutionary
stage and thus to the initial conditions. Whether the variability to
the observed emission will continue on longer timescales (i.e
$10^5\;$yrs) is not clear as it depends on the postshock gas
pressure. The luminosity of the cloud material follows the morphology
of the winds, which suggests the magnitude and variability of this
emission is related to the activities of the winds.

\begin{figure*}
\begin{center}
    \begin{tabular}{ccc}
      \resizebox{50mm}{!}{{\includegraphics{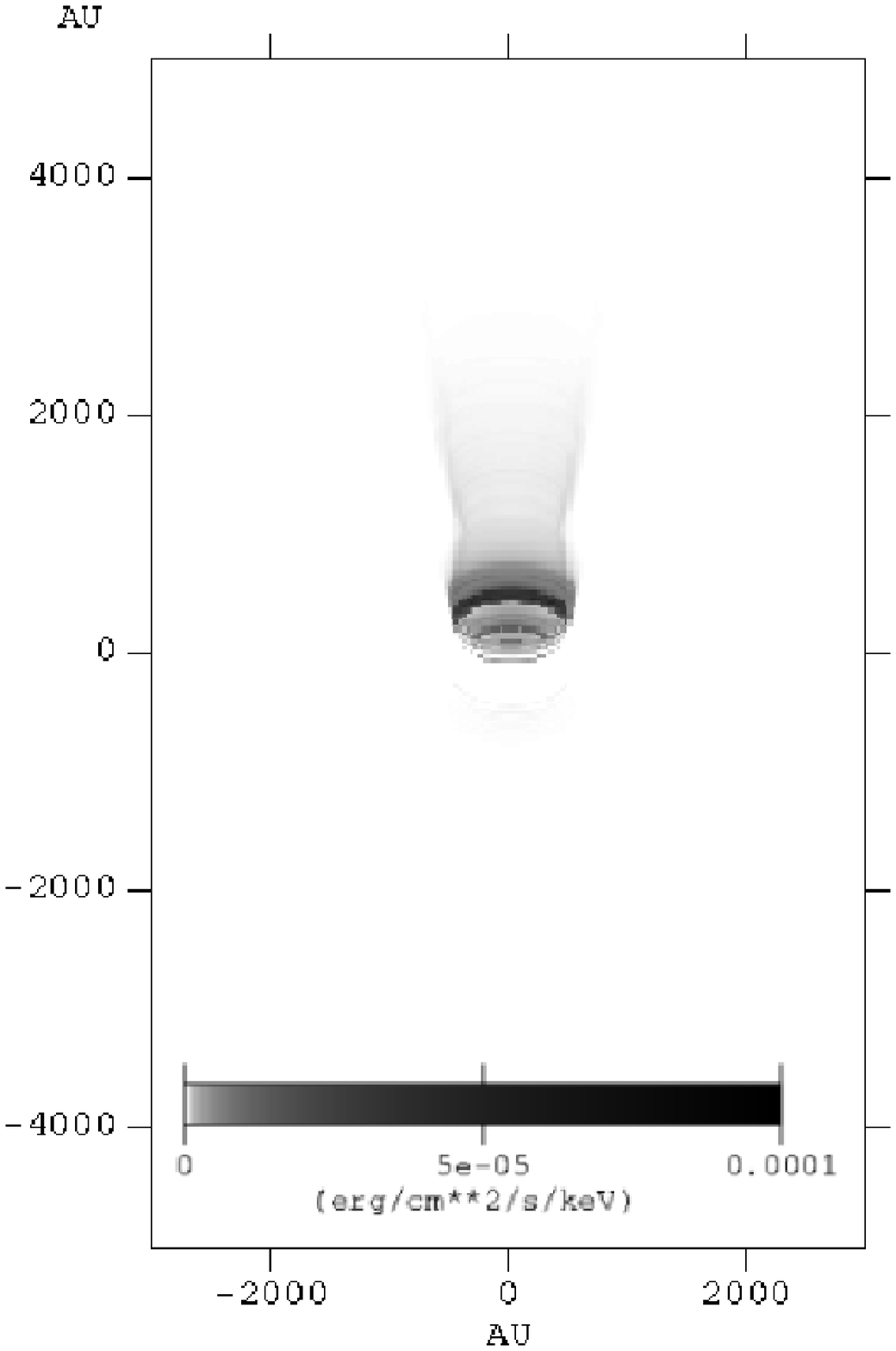}}} &
      \resizebox{50mm}{!}{{\includegraphics{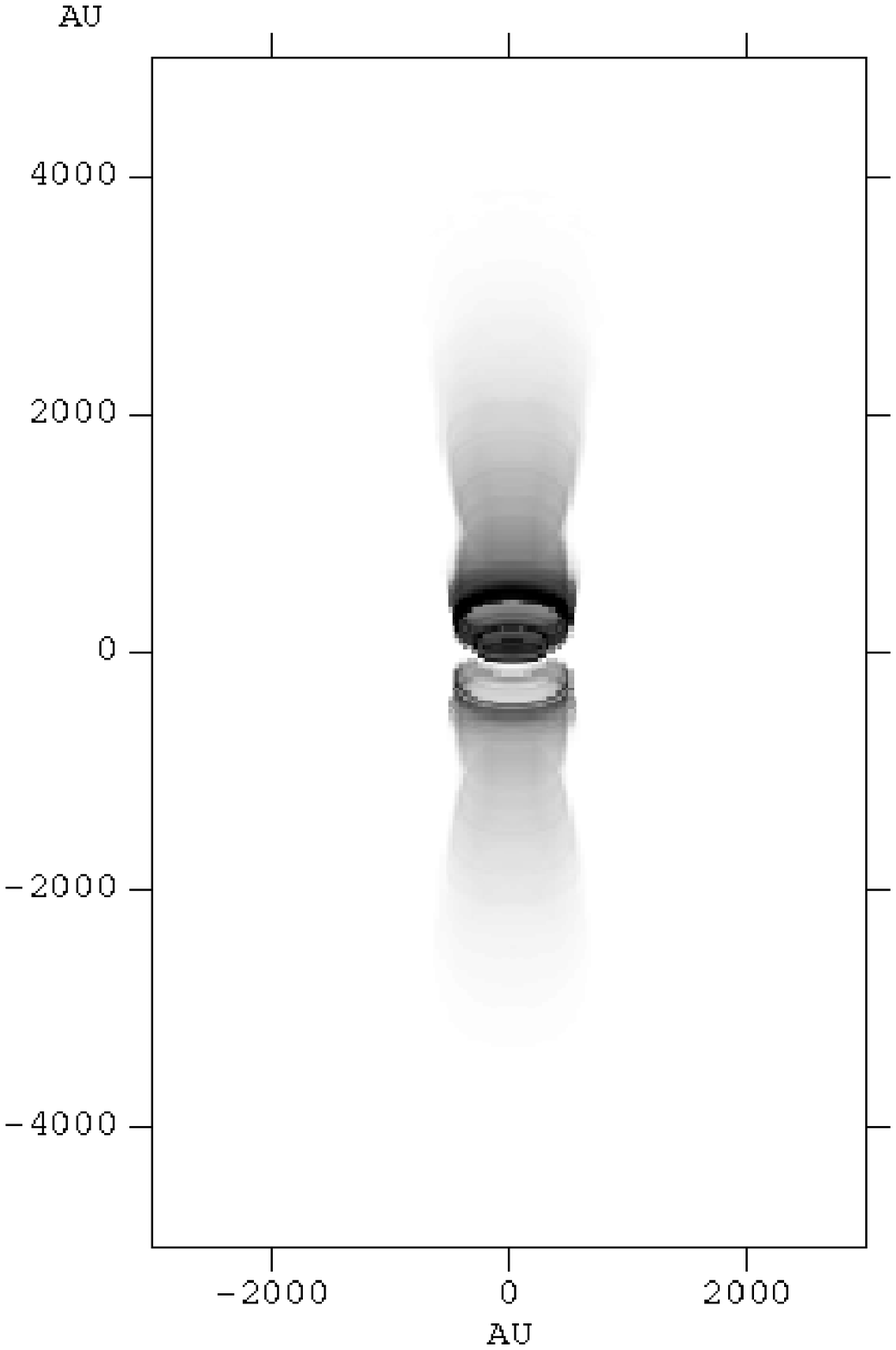}}} &
      \resizebox{50mm}{!}{{\includegraphics{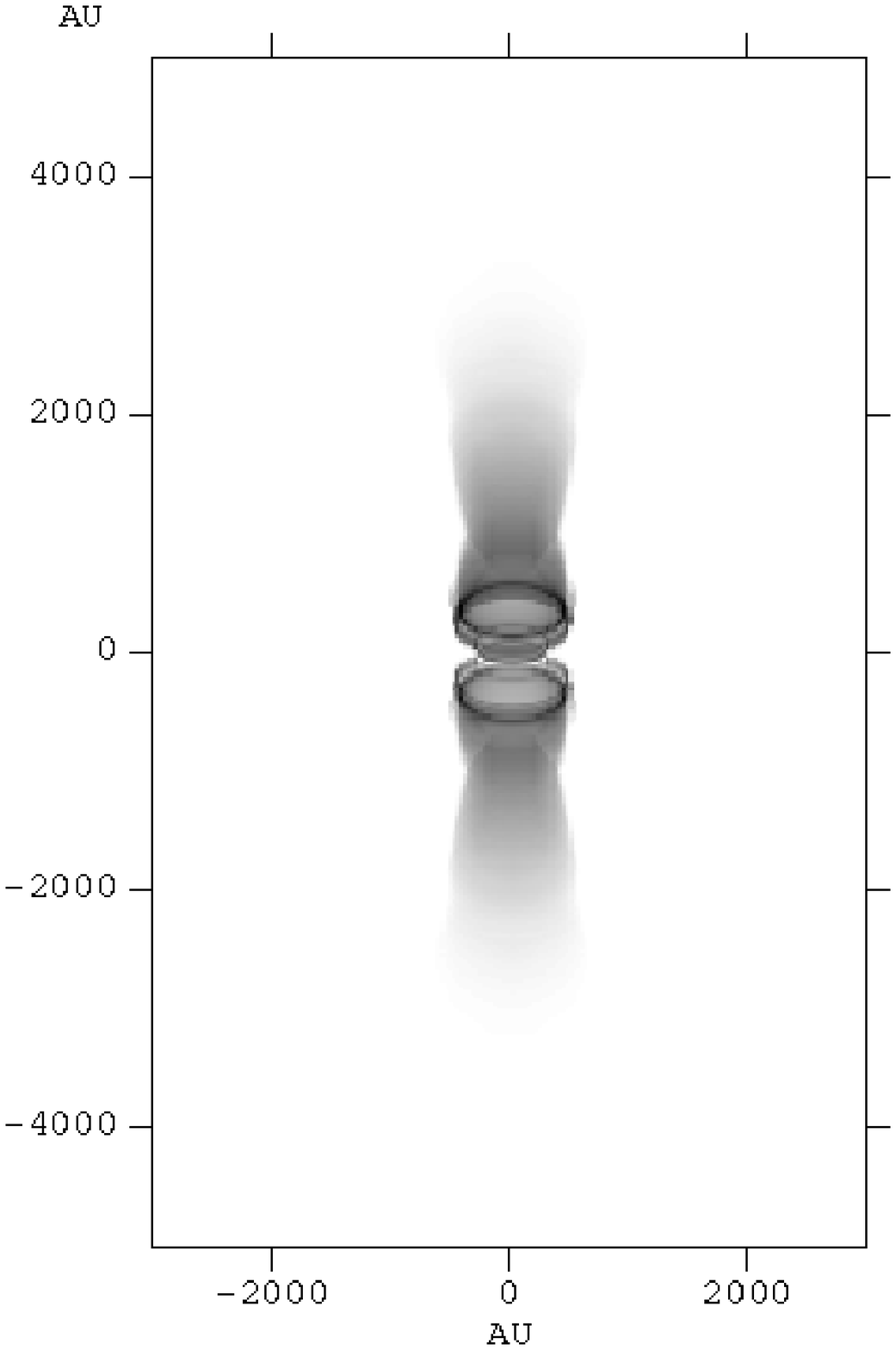}}} \\
    \end{tabular}
    \caption[]{Broadband images of the attenuated X-ray emission from
    model R1 at $t=1000\;$yr, an inclination angle of $i=60^{\circ}$,
    a distance of 1 kpc, and in the energy bands 1-2 (left panel), 2-4
    (middle panel) and 4-10 keV (right panel).}
    \label{fig:broadbandimages}
\end{center}
\end{figure*}

The visual extinction to a deeply embedded star relates to the column
density along that line-of-sight.  To allow comparison between our
calculated column densities and observationally determined visual
exctinctions, $A_{\rm v}$'s, we note the simple conversion factor
$A_{\rm v}=N_{\rm H}/(1.9\times10^{21}\;\cm^{-2} )\;{\rm mag}$
\citep[e.g.][]{Cox:2000}.  One method of observationally determining
the optical depth to the MYSO is to use the 9.7 $\mu$m Silicate
feature \citep{Crapsi:2008}. For the simulations, a good approximation
is the column density to the star, $N_{\rm H-star}$ (i.e. the grid
origin). Comparing this value to the emission weighted column (EWC),
$N_{\rm EWC}= \Sigma N_{\rm H c} L_{\rm int T c} / \Sigma L_{\rm int T
c} $ \citep[e.g.][]{Parkin:2008}, where the index $c$ indicates the
column and luminosity from each cell on the hydrodynamic grid and the
summation is over all cells, shows that the column density to regions
with large intrinsic X-ray emission is significantly higher than that
to the star (Table~\ref{tab:xrays}). This is mainly due to
lines-of-sight to the X-ray emitting plasma in the receding lobe
having longer path lengths and/or passing through denser material. For
an inclination angle of $i=60^{\circ}$, the majority of the absorbing
column is through the first $\sim2000\;$au of cloud gas beyond the
cavity wall with a large fraction due to the layer of compressed cloud
material which lines the cavity wall.

\subsubsection{Inclination angle dependence of the observed X-ray emission}

The attenuation of the emission is viewing angle dependent. When the
inclination angle is $i=30^{\circ}$ the majority of lines-of-sight to
X-ray emitting plasma avoid the densest cloud material close to the
equatorial plane. This results in a lower $N_{\rm H-star}$ and
significantly brighter spectrum at $E \ltsimm 4\;$keV than at
$i=60^{\circ}$ (Fig.~\ref{fig:incspectra}). At $E \gtsimm 4\;$keV the
harder stellar wind component dominates the spectrum. The disk wind
and heated cloud material contribute the majority of the X-ray
emission, with little difference between the two ($L_{\rm att_{D}} =
1.24\times10^{31}\ergps$ and $L_{\rm att_{C}} =
1.61\times10^{31}\ergps$, respectively). The total attenuated
luminosity ($L_{\rm att_ {T}} = 2.88\times10^{31}\;\ergps$), and the
integrated 0.1-10 keV count rate ($\eta\simeq 30\;{\rm ks}^{-1}$) are
two orders of magnitude higher than observationally detected count
rates.

When $i=45^{\circ}$, the majority of the X-ray emitting plasma is
viewed through the cloud material. The column densities have increased
relative to $i=30^{\circ}$ as now the line-of-sight to the star and to
the regions with highest intrinsic emission must pass through sections
of higher density, compressed cloud material. Examining the spectrum
(Fig.~\ref{fig:incspectra}) the turnover energy is now at $E\simeq
1.5\;$keV. This is an increase relative to the $i=30^{\circ}$ case
which illustrates the higher attenuation to the X-ray emission from
disk wind and cloud material. At energies $E\gtsimm4\;$keV the
spectrum has not changed significantly. $L_{\rm att_{T}}$ and $\eta$
are now more in line with X-ray observations of massive star forming
regions
\citep[e.g.][]{Kohno:2002,Preibisch:2002,Giardino:2004}. Increasing
the inclination angle further to $i=60^{\circ}$ results in a slight
rise in $N_{\rm H -star}$, $N_{\rm EWC}$, and the turnover energy of
the spectrum. Similarly, $L_{\rm att_{T}}$ and $\eta$ decrease.

The maximum column densities are attained when viewing the system
edge-on (i.e. $i=90^{\circ}$). It is interesting to note that the
column to the star is now greater than the EWC, which indicates that
for this viewing angle the column to the spatially extended (few
thousand au) distribution of X-ray emitting plasma is lower than that
to the star. Unlike the lower inclination angle calculations, emission
below $E\ltsimm 2\;$keV is now observed from both lobes of the
cavity. Also, the stellar wind material, which has the lowest
intrinsic luminosity, now provides the largest contribution to $L_{\rm
att_{T}}$. The spectrum above $E\sim2\;$keV remains roughly
similar. We note that for inclination angles where the line-of-sight
to the X-ray emitting plasma passes through the cloud material
(i.e. $i\gtsimm45^{\circ}$ when $\omega \simeq 30^{\circ}$), emission
above $E\sim4\;$keV is mainly contributed by the shocked stellar wind.

The inclination angle dependence of the observed emission shows that
there is a narrow range of angles where the count rate exceeds the
\textit{Chandra} X-ray detections. 

\vspace{-3mm}
\subsubsection{Synthetic X-ray images}

The raytracing code can produce synthetic broadband X-ray images which
allow us to identify where different energy X-rays preferentially
originate from. For simulation R1 at $i=60^{\circ}$, we see that the
observable X-ray emission in the 1-2, 2-4, and 4-10 keV bands
originates from similar regions of the cavity
(Fig.~\ref{fig:broadbandimages}). A peak in the intensity in the three
bands originates near the reverse shock, and is mainly generated by
shocked stellar and disk wind material
(Fig.~\ref{fig:incspectra}). Emission extending in the polar direction
traces the boundary between the hot shocked stellar wind, and the
cooler, higher density disk wind.

Examining Fig.~\ref{fig:cavevo} shows that the region of the cavity
close to the pole is filled by hot shocked stellar wind. This gas
emits the majority of the emission above $E\simeq4\;$keV
(Fig.~\ref{fig:incspectra}). The tower of emission visible in the 4-10
keV image in Fig.~\ref{fig:broadbandimages} therefore traces the
stellar wind, with the shocked cloud emission in this energy band
originating from the base of the cavity wall. The peak in intensity in
all three bands occurs at the waist of the reverse shock, where there
is a combined maximum in temperature and density. Examining the
spectrum in the $E=2-4\;$keV energy range shows that this emission is
from shocked disk wind and cloud material, which suggests it
originates from the $T\sim10^{6.5}\;$K gas at $z=r\simeq
0.5\times10^{16}\;\cm$. Interestingly, both the preceding and receding
lobes are observable in the 2-4 and 4-10 keV bands, which signifies the
lessening influence of attenuation as the photon energy increases.

Importantly the spatial extents of the emission in all three energy
bands (1-2, 2-4, and 4-10 keV) are just below the resolution limit of
\textit{Chandra} ($\sim0.5''$) and so this model is consistent with
the unresolved nature of real MYSOs. However, this finding is not true
for all of our models which allows us to place constraints on some of
the key model parameters.

\vspace{-3mm}
\subsection{Variation with mass inflow/outflow rates}
\label{subsec:massflowrates}

The velocities of the inflowing and outflowing gas in the simulations
are supersonic. As such, the morphology of the evolving cavity is
strongly dependent on the ratio of the inflow and outflow ram
pressure, which is directly proportional to the mass flow rates. There
are a multitude of possible models with identical ram pressure ratios,
but producing vastly different observable X-ray emission
characteristics. Since optically thin bremsstrahlung emission has a
density squared dependence (the emissivity $\propto n^{2}$), an
increase in the mass-loss rates can have a considerable impact on the
observed flux. However, an increase in the emission from cloud
material due to an increase in $\dot{M}_{\rm infall}$ may be countered
by an accompanying increase in attenuation. The degree of absorption
is dependent on the column density, which is largely due to the cold
cloud material, and therefore is directly proportional to the mass
infall rate ($N_{\rm H}\propto\dot{M}_{\rm infall}$).

Decreasing the mass infall rate reduces the ram pressure of the
infall. In model R2 the mass infall rate is halved to $\dot{M}_{\rm
infall}=1\times10^{-4}\Msolpyr$. This results in less resistance to
the outflows and subsequent expansion of the cavity. The cavity wall
is hence wider at the base at an equivalent time. Also, the reverse
shock is positioned at a greater distance from the star, and the
enclosed region appears more elongated in the $z$-direction. Apart
from the afore-mentioned differences we note that simulations R2 and
R1 show similar evolution. As expected, $N_{\rm H-star}$ and $N_{\rm
EWC}$ decrease by roughly a factor of two, consistent with the
decrease in $\dot{M}_{\rm infall}$. There is a slight increase in
$L_{\rm int_{T}}$ but $L_{\rm att_{T}}$ and $\eta$ increase by factors
of 2 and 3.5 respectively. Compared to model R1, the spectrum for
model R2 (Fig.~\ref{fig:spectra}) shows higher flux at $E\ltsimm
4\;$keV and a softer spectrum above this energy which is linked to the
differences in the position of the reverse shock.

\begin{figure}
\begin{center}
    \begin{tabular}{l}
      \resizebox{60mm}{!}{{\includegraphics{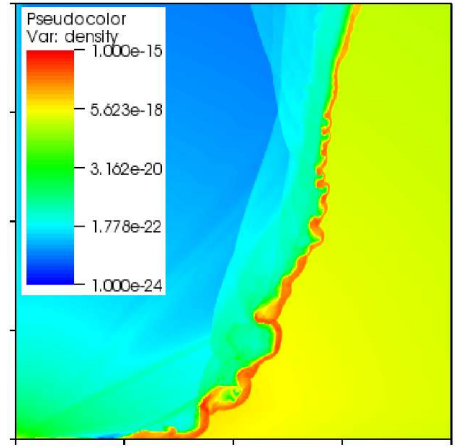}}} \\
    \end{tabular}
    \caption[]{Density snapshot taken from model R4 at
    $t=2000\;$yr. The tick marks on the axis correspond to a distance
    of $10^{16}\;$cm.}
    \label{fig:instabilities}
\end{center}
\end{figure}

\begin{figure}
\begin{center}
    \begin{tabular}{ll}
      \resizebox{40mm}{!}{{\includegraphics{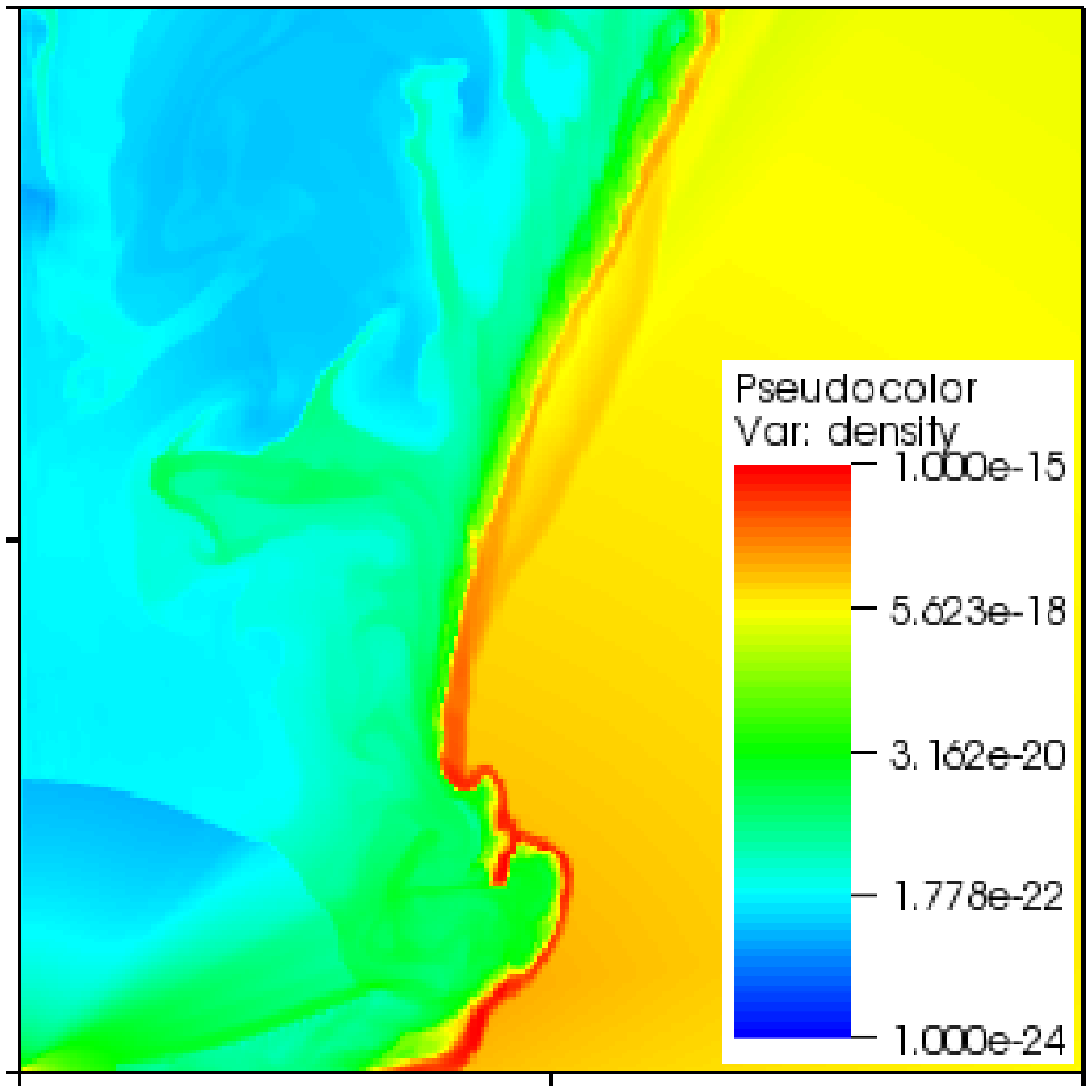}}} &
      \resizebox{40mm}{!}{{\includegraphics{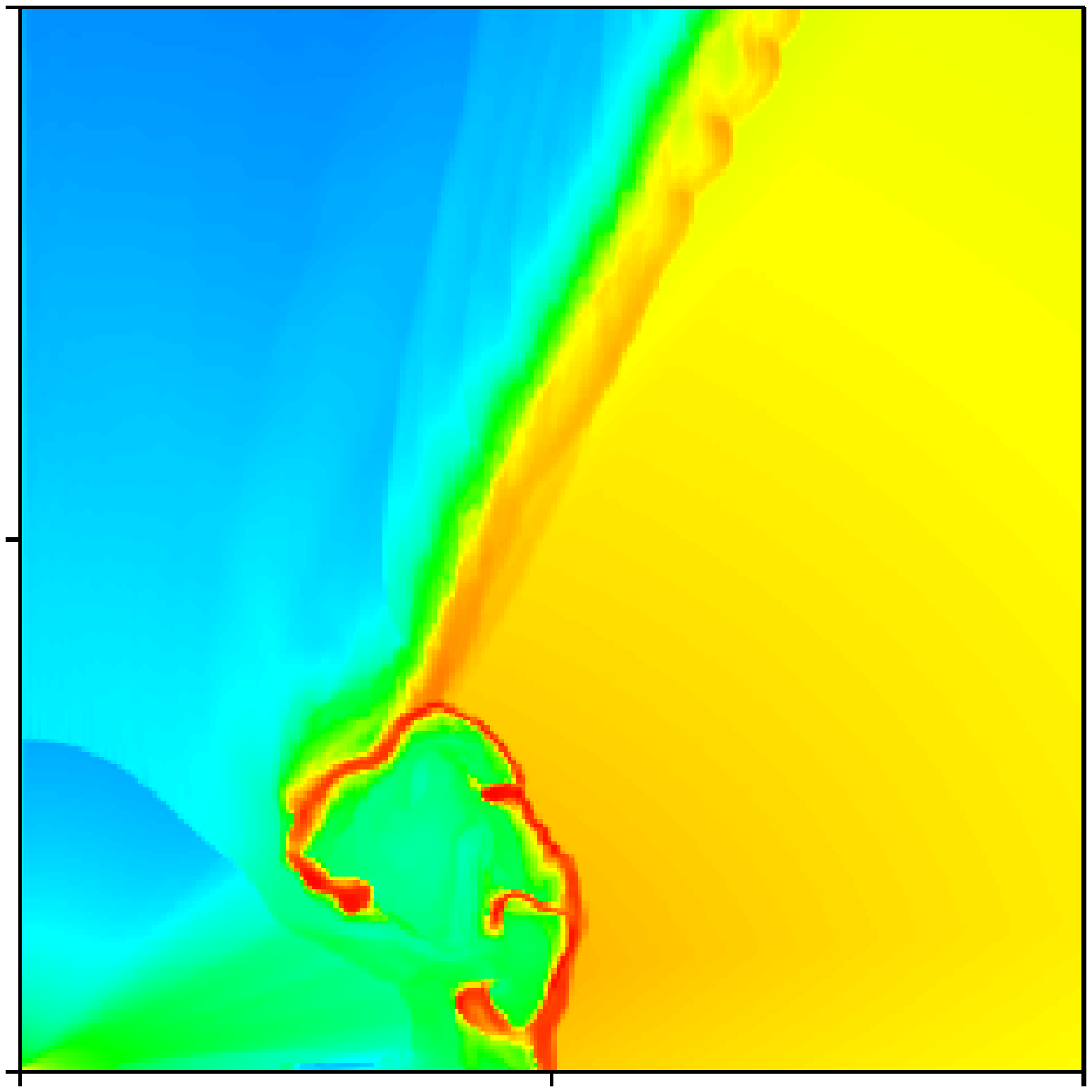}}} \\
    \end{tabular}
    \caption[]{Density snapshots taken from model R7 at $t=$1700
      (left) and 2000 yrs (right). The tick marks on the axis
      correspond to a distance of $10^{16}\;$cm.}
    \label{fig:hole}
\end{center}
\end{figure}

If the mass infall rate is increased the ram pressure of the infalling
material increases in direct proportion and the outflowing winds
encounter more resistance to their expansion. In model R3 the reverse
shock is pushed closer to the star by the increased mass infall rate,
and the cavity now closes up over the star. At $t=500\;$yr the infall
has begun to obstruct some of the outflow, and by $t=900\;$yrs the
outflow has been fully constricted. Table~\ref{tab:xrays} contains
X-ray luminosities at $t=500\;$yr, before the cavity closes up (once
closed the X-ray emission in our model becomes zero). Examining
Table~\ref{tab:xrays}, we find that $N_{\rm H-star}$ and $N_{\rm EWC}$
have increased, and the $L_{\rm att}$'s and $\eta$'s have decreased
for R3 relative to R1. Note that $L_{\rm int C}$, increases
considerably with $\dot{M}_{\rm infall}$, as the density, and thus the
emission measure of this material, increases. However, the increasing
attenuation with $\dot{M}_{\rm infall}$ removes more and more of the
soft X-ray emission, so that the observed emission becomes
increasingly hard (see Fig.~\ref{fig:spectra}). Model R3 shows that
there is a delicate hydrodynamic balance between the infall and
outflow, and large contrasts between these parameters can result in
vastly different cavity evolution.

Confinement of a wind was discussed in the context of T-Tauri stars by
\cite{Chevalier:1983}, where the infall ram pressure can suppress the
expansion of a wind into the surrounding envelope. We have a similar
situation, albeit more complicated by the anisotropic outflow. The
wind at the base of the cavity may have a higher ram pressure than the
stellar wind towards the pole of the cavity (as in model R3), which
may ultimately lead to confinement at the poles but expansion at the
base. Note also that the simulations we have performed only consider
hydrodynamics, whereas a number of physical processes may in reality
prevent a cavity from closing up. Magnetic fields may play an
important role in controlling the dynamics of the infalling envelope
\citep[e.g.][]{Banerjee:2007}, or the radiation pressure may halt the
inflow \citep[i.e.][]{Krumholz:2005}. However, it is clear that within
the context of our model, the parameters for a slowly expanding cavity
lie within a narrow range.

In model R4 the mass-loss rate of the stellar wind is increased to
$1\times10^{-6}\Msolpyr$. The boundary between the inflowing and
outflowing gas is likely to be subject to dynamical shear
instabilities. Unlike in R1, such instabilities can be seen forming in
the lower part of the cavity wall after $t\sim500\;$ yr, with
prominent structure visible at $t=2000\;$yrs
(Fig.~\ref{fig:instabilities}). A common feature in the simulations
where these instabilities are noticeable (R4, R5, R9, R14, R15, and
R18) is that the disk wind is compressed into a thin layer along the
cavity wall. The timescale for KH instabilities to grow is $\propto
(\rho_{1} + \rho_{2})/ \sqrt{\rho_{1} \rho_{2}}$ , and the enhanced
disk wind density along the cavity wall results in an increased growth
rate for these instabilities. In addition to this, in the
afore-mentioned simulations the cavity wall becomes curved at the base
due to the action of the winds (except in R18 where the cavity is
initially curved). As such, the shocked winds have a small grazing
angle against the base of the cavity wall and the velocity difference
across the shear boundary is higher. These instabilities cause the
initially smooth cavity wall to become deformed, and in the more
severe cases these deformations grow into jagged features. As these
features protrude into the winds clumps of cloud material are stripped
away and then ablated. In model R1, we see disk wind material mixing
into stellar material (see Fig.~\ref{fig:cavevo}), while in model R4
(Fig.~\ref{fig:instabilities}) we see cloud material mixing into disk
wind material. Such mixing can cause the disk wind to become more
confined to the lower regions of the simulation domain, and no longer
border the entire cavity wall. Also of note in model R4 is that the
wind-driven expansion of the cavity opposes the infall and it is
conceivable that at $t>2000\;$yr the flow of cloud material onto the
disk plane will be halted and the envelope will be completely
destroyed by the winds. The increase in $\dot{M}_{\rm S}$ results in
an increase in the $L_{\rm att}$'s and $\eta$.

In model R5 the ratio of the inflow/outflow ram pressures is the same
as in R4, but $\dot{M}_{\rm infall}$, $\dot{M}_{\rm S}$, and
$\dot{M}_{\rm D}$ are all ten times lower. Nevertheless, the evolution
of the cavity and the position and shape of the reverse shock are
almost identical between R4 and R5, as one would expect. The intrinsic
emission decreases by a factor of 100, as expected ($EM \propto
\dot{M}^{2}$). However, $\eta$ only decreases by a factor of $\sim 2$,
since the attenuation is also vastly reduced. Comparing the spectra
(Fig.~\ref{fig:spectra}) we see that in the 4-10 keV band the spectral
slope is almost identical, yet for R5 the count rate in this band is
roughly a factor of 100 lower. Therefore, although count rates
consistent with observations can still be attained when the mass flow
rates are scaled down, the much softer observed spectrum places
constraints on the outflow, and consequently the inflow, parameters.

In R6 the mass-loss rate of the disk wind is increased by a factor of
3 from R1 to $\dot{M}_{\rm D}=3\times10^{-6}\Msolpyr$. If
$\dot{M}_{\rm infall}$ is kept at $2\times10^{-4}\Msolpyr$ the disk
wind drives a hole into the base of the cavity wall. To prevent this
occurence $\dot{M}_{\rm infall}$ is increased to
$1\times10^{-3}\Msolpyr$. The resulting cavity evolution is somewhat
similar to that of R1. However, the fact that $\dot{M}_{\rm infall}$
has been increased to stabilise the cavity evolution brings an
accompanying increase in the column densities so that the attenuated
luminosity is actually lower than that from model R1. Compared to
model R3 (which has identical parameters except for a lower value of
$\dot{M}_{\rm D}$) we see that the higher value of $\dot{M}_{\rm D}$
prevents the infall from totally overwhelming the outflow.

\begin{figure}
\begin{center}
    \begin{tabular}{ll}
      \resizebox{40mm}{!}{{\includegraphics{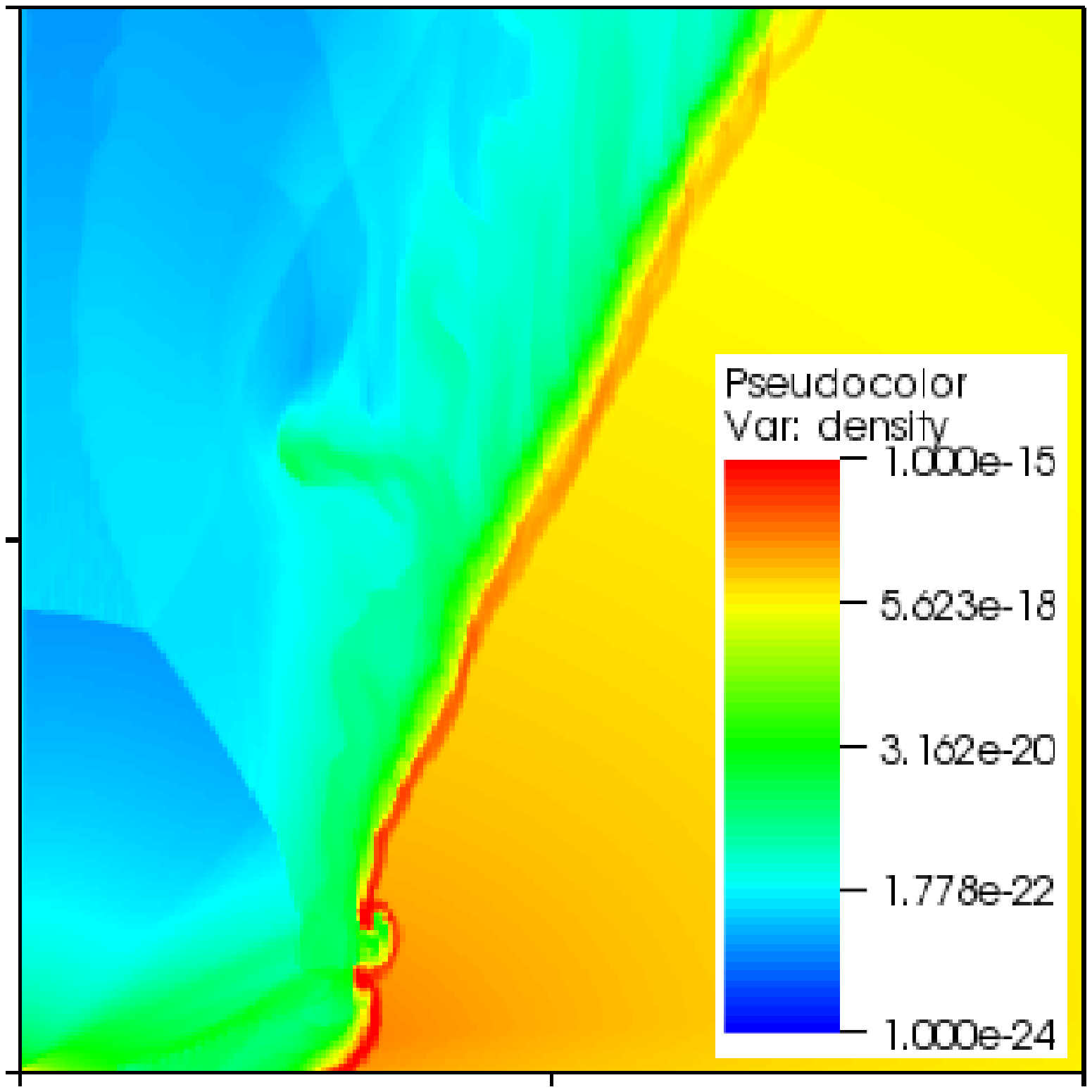}}} &
      \resizebox{40mm}{!}{{\includegraphics{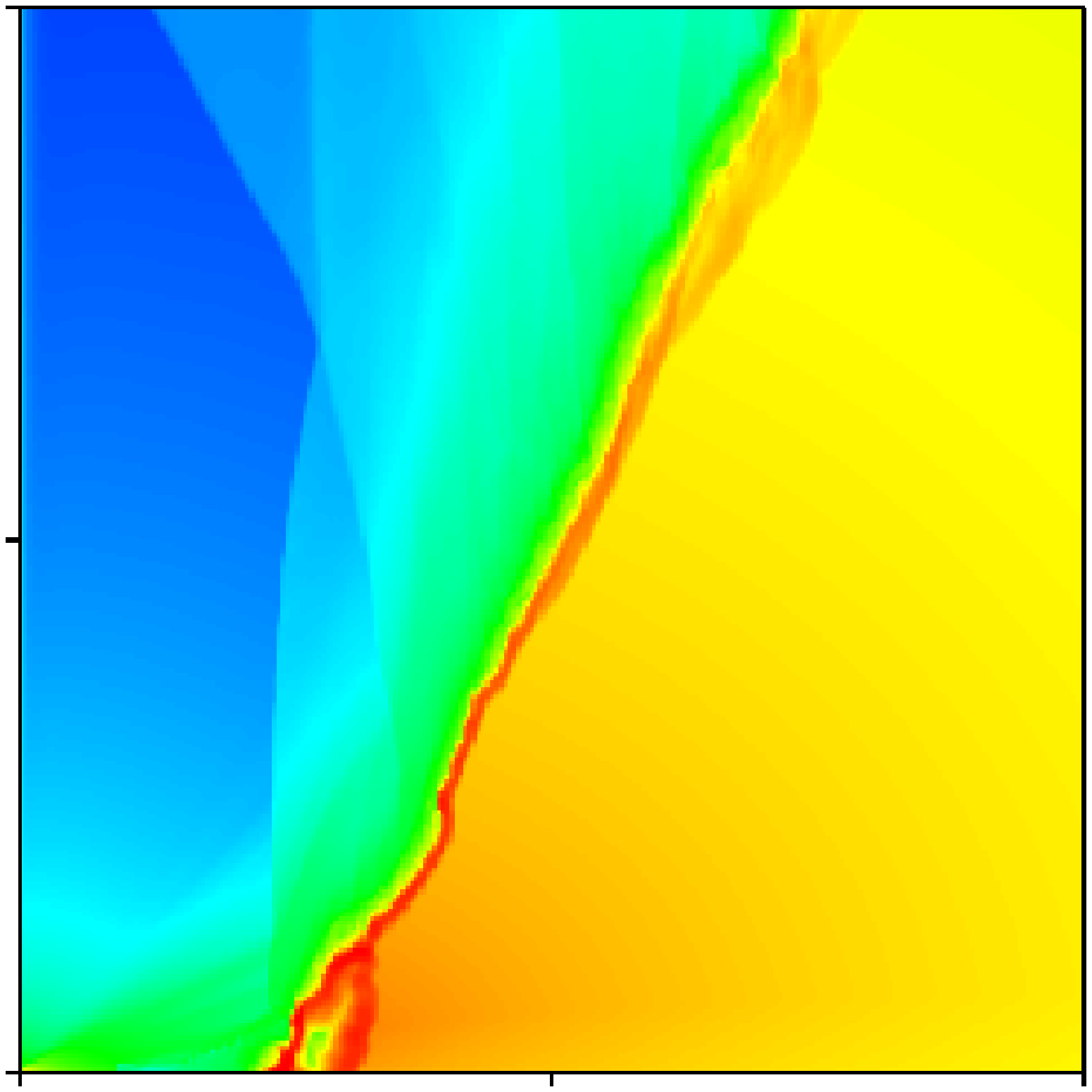}}} \\
    \end{tabular}
    \caption[]{Density snapshots taken from model R8 at
      $t=600\;$(left) and 900 yrs (right). The tick marks on the axis
      correspond to a distance of $10^{16}\;$cm.}
    \label{fig:smallhole}
\end{center}
\end{figure}

Reducing the mass-loss rate of the stellar wind (R7) leads to major
differences in the cavity evolution. The pressure applied to the
cavity wall by the winds as a function of polar angle now differs more
than in model R1 (there is now a factor of 20 variation in
$\dot{M}_{\theta}$). Whereas the stellar wind previously had
sufficient ram pressure to hold back the infalling cloud material, the
cavity now begins to close up above the star. At $t\simeq1500\;$yrs
the disk wind begins to cut into the base of the cavity wall, and the
bubble formed continues to increase in size for the duration of the
simulation (Fig.~\ref{fig:hole}). By $t=2000\;$yrs a limb of cloud
material is extending towards the star. Further analysis shows that by
$t=2300\;$yrs this material breaks away from the cavity wall and
obstructs the outflowing winds close to the star.

\begin{table*}
  \centering 
  \caption[]{Column densities (columns 3 and 4), intrinsic (int) and
  attenuated (att) 0.1-10 keV X-ray luminosities (columns 5-8 and
  9-12, respectively), and integrated count rates (column 13) from
  raytraced emission calculations. The stellar wind (S), disc wind
  (D), and cloud material (C) contributions to the total (T)
  luminosity are shown. The integrated 0.1-10 keV count rate, $\eta$,
  was calculated using the \textit{Chandra} effective area (see
  \S~\ref{subsec:xrayemission}). Luminosities are in
  $10^{31}\;\ergps$, and count rates are in ks$^{-1}$. $N_{\rm
  H-star}$ and $N_{\rm EWC}$ are the column density measured to the
  star and the emission weighted column respectively, both of which
  are in $10^{23}\;{\rm cm^{-2}}$. The raytracing calculations were
  performed with an inclination angle of $i=60^{\circ}$, except
  R1$_{30}$, R1$_{45}$, and R1$_{90}$ which were calculated using
  inclination angles of 0, 30, 45, and $90^{\circ}$ respectively. All
  calculations were performed on simulations at $t=1000\;$yr, except
  R1$_{\rm long}$ and R3 which were calculated at 5000 and 500 yrs,
  respectively.}  \centering
\begin{tabular}{llccccccccccc}
\hline
Model & Comment & $N_{\rm H-star}$ & $N_{\rm EWC}$ & ${\rm L_{int_{T}}}$ & ${\rm L_{int_{S}}}$ & ${\rm L_{int_{D}}}$ & ${\rm L_{int_{C}}}$ & ${\rm L_{att_{T}}}$ & ${\rm L_{att_{S}}}$ & ${\rm L_{att_{D}}}$ & ${\rm L_{att_{C}}}$ & $\eta$ \\
 & & & & & & & & & & & & \\ 
\hline
R1 & & 2.47 & 6.09 & 370 & 0.500 & 43.1 & 330 & 0.019 & 0.007 & 0.004 & 0.008 & 0.11 \\
R1$_{30}$ & &  0.44 & 5.42 & 370 & 0.500 & 43.1 & 330 & 2.88 & 0.035 & 1.24 & 1.61 & 30 \\
R1$_{45}$ & & 3.11 & 6.21 & 370 & 0.500 & 43.1 & 330 & 0.040 & 0.008 & 0.012 & 0.020 &  0.35 \\
R1$_{90}$ & & 9.66 & 7.03 & 370 & 0.500 & 43.1 & 330 & 0.012 & 0.006 & 0.002 & 0.004 &  0.05 \\
\hline
R1$_{\rm long}$ & longer run & 2.40 & 5.54 & 186 & 0.221 & 26.1 & 157 & 0.007 & 0.002 & 0.004 & 0.001 & 0.05 \\
\hline
R2 & $\dot{M}_{\rm infall} \downarrow$ & 0.99 & 3.19 & 436 & 0.313 & 32.2 & 404 & 0.040 & 0.005 & 0.021 & 0.014 & 0.37 \\
R3 & $\dot{M}_{\rm infall} \uparrow$ & 16.7 & 46.4 & 3070 & 0.596 & 27.7 & 3050 & 0.014 & 0.002 & 0.005 & 0.007 & 0.02 \\
R4 & $\dot{M}_{\rm S} \uparrow$ & 1.92 & 3.46 & 220 & 1.39 & 33.4 & 185 & 0.071 & 0.043 & 0.015 & 0.012 & 0.41 \\
R5 & R4 with $\dot{M}$'s $\downarrow$ & 0.25 & 0.37 & 2.48& 0.11 & 0.22 & 2.16 & 0.016 & 0.002 & 0.006 & 0.008 & 0.23 \\
R6 & $\dot{M}_{\rm infall} \uparrow$, $\dot{M}_{\rm D} \uparrow$ & 16.9 & 46.6 & 6120 & 5.27 & 395 & 5720 & 0.008 & 0.007 &
$7\times10^{-4}$ & $5\times10^{-4}$ & 0.02 \\
R7 & $\dot{M}_{\rm S} \downarrow$ & 2.47 & 6.39 & 390 & 0.470 & 44.1 & 345 & 0.009 & 0.005 & 0.002 & 0.002 & 0.05 \\
\hline
R8 & $v_{\rm S} \uparrow$ & 2.48 & 7.38 & 569 & 0.142 & 33.8 & 535 & 0.049 & 0.005 & 0.003 & 0.011 & 0.26 \\
R9 & $v_{\rm S} \downarrow$, $\dot{M}_{\rm S} \uparrow$ & 2.10 & 4.40 & 215 & 4.49 & 36.8 & 174 & 0.042 & 0.027 & 0.008 & 0.007 & 0.28 \\
R10 & B star & 0.38 & 0.96 & 13.5 & 0.066 & 2.26 & 11.2 &  0.004 & $4\times10^{-4}$ & $7\times10^{-4}$ & 0.003 & 0.06 \\
\hline
R11 & $\theta_{\rm S}=70^{\circ}$ & 2.83 & 7.42 & 438 & 1.64 & 31.5 & 405 &  0.019 & 0.013 & 0.003 & 0.003 & 0.08 \\
R12 & $\theta_{\rm S} = 85^{\circ}$, $\theta_{\rm D}=89^{\circ}$ & 3.00 & 9.15 & 526 & 2.41 & 6.93 & 516 & 0.030 & 0.017 & 0.002 & 0.012 & 0.12\\
R13 & no disk wind & 2.93 & 9.52 & 456 & 2.22 & $-$ & 454 & 0.060 & 0.032 & $-$ & 0.029 & 0.20 \\
\hline
R14 & $\omega = 5^{\circ}$ & 2.41 & 6.66 & 318 & 0.253 & 27.8 & 290 & 0.051 & 0.011 & 0.016 & 0.024 & 0.30 \\
R15 & $\omega = 10^{\circ}$ & 2.82 & 7.22 & 483 & 0.577 & 54.5 & 4.28 & 0.009 & 0.006 & 0.002 & 0.002 & 0.05 \\
R16 & $\omega = 45^{\circ}$ & 2.14 & 6.40 & 219 & 0.219 & 23.8 & 195 & 0.018 & 0.004 & 0.012 & 0.002 & 0.12\\
R17 & $\omega = 60^{\circ}$ & 1.21 & 3.36 & 27.5 & 0.125 & 11.9 & 15.4 & 0.003 & $2\times10^{-5}$ &$3\times10^{-4}$ & 0.003 & 0.05 \\
\hline
R18 & $\beta = 2$ & 1.36 & 3.40 & 341 & 1.11 & 30.3 & 310 & 0.099 & 0.036 & 0.015 & 0.048 & 0.58 \\
\hline
R19 & $r_{\rm c} \downarrow$ & 2.86 & 7.42 & 754 & 0.378 & 45.2 & 708 & 0.025 & 0.006 & 0.009 & 0.010 & 0.14 \\
R20 & $r_{\rm c} \uparrow$ & 2.39 & 6.21 & 148 & 0.158 & 11.7 & 136 & 0.079 & 0.008 & 0.059 & 0.012 & 0.39 \\
\hline
\label{tab:xrays}
\end{tabular}
\end{table*}

\vspace{-3mm}
\subsection{Variation with wind speeds}

The preshock velocity dictates the postshock gas temperature ($T
\propto v^{2}$). Hotter gas emits higher energy X-rays, and an
increase or reduction in the wind speeds will affect the hardness of
the resulting spectrum.

In model R8 the stellar wind velocity is increased to $v_{\rm
S}=3000\;{\rm km s^{-1}}$. The shocked stellar wind now reaches
$T\gtsimm10^{8}\;$K. The equatorial flow appears to be more powerful
and drives a small, short-lived hole into the base of the cavity wall
(Fig.~\ref{fig:smallhole}). This is unexpected as it is the ram
pressure of the stellar wind which has been increased. However, it
seems that the total pressure near the equatorial plane is enhanced by
an increase in the thermal pressure adjacent to the reverse
shock. Subsequently, the cloud material then collapses (see
Fig.~\ref{fig:smallhole}) and fills the hole, after which the cavity
settles into a steady state similar to that of model R1. The higher
ram pressure of the stellar wind increases the maximum distance of the
reverse shock from the star, and its amplitude of
oscillation. Increasing the stellar wind velocity causes the spectral
hardness (Fig.~\ref{fig:spectra}) and the \textit{Chandra} count rate,
$\eta$, to increase. Interestingly, the X-ray emission from the disk
wind and cloud material is now dominant at all energies. Whereas
$L_{\rm att_{S}}$ and $L_{\rm att_{C}}$ were approximately equal in
model R1, $L_{\rm att_{C}}$ is twice as high as $L_{\rm att_{S}}$ in
model R8. This is likely a result of increased turbulent heating by
the stellar wind. A decrease in $L_{\rm int_{S}}$ occurs because the
reverse shock is further from the star, leading to a lower postshock
density, and thus a lower emission measure.

\begin{figure}
\begin{center}
    \begin{tabular}{c}
      \resizebox{50mm}{!}{{\includegraphics{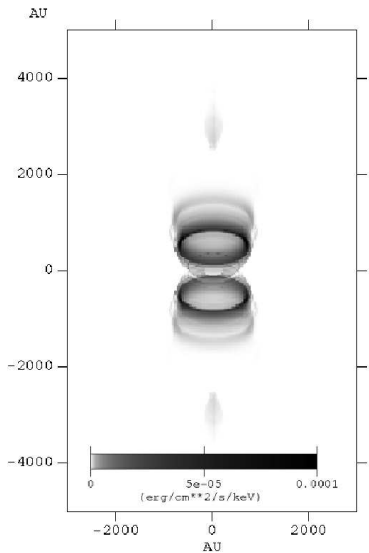}}} \\
    \end{tabular}
    \caption[]{Broadband image of the attenuated X-ray emission for
    model R9 at an inclination angle of $i=60^{\circ}$, a distance of
    1 kpc, and in the 4-10 keV energy band.}
    \label{fig:cavrun28_bbimage}
\end{center}
\end{figure}

Recent theoretical models of stellar evolution by \cite{Hosokawa:2009}
predict that for high accretion rates ($\sim 10^{-3}\Msolpyr$) the
protostellar radius becomes very large. This results in a lower
surface gravity for the protostar and consequently the stellar wind
will be more like that of a supergiant (i.e. $v_{\rm S}\ltsimm
1000\;$\kmps) than a main sequence O-star. Model R9 explores this
scenario, where $\dot{M}_{\rm S}=\dot{M}_{\rm D}=10^{-6}\Msolpyr$ and
$v_{\rm S}=1000\kmps$. With a greater ram pressure in the stellar wind
compared to model R1, the cavity does not initially narrow at the top
of the grid. The postshock stellar wind has a lower temperature
($T\sim10^{6.5}\;$K) because of the lower preshock velocity. The
position of the reverse shock is much further from the star
($z\sim4\times10^{16}\cm$), and this distance increases as the
simulation progresses. The column densities are lower
(Table~\ref{tab:xrays}), likely due to the greater height of the
reverse shock (and thus the postshock gas) above the disk plane, and
thus the reduced path length through cloud material for
lines-of-sight. The spectrum is very different from that of R1 and is
clearly a lot softer (Fig.~\ref{fig:spectra}). The stellar wind
component now dominates the observed emission at $E\sim1\;$keV as the
denser stellar wind increases the emission measure. The synthetic
X-ray images show the approaching lobe to be fully illuminated by
X-ray emission, unlike the narrow tower seen for R1 in
Fig.~\ref{fig:broadbandimages}, with the intensity peaks in the 1-2,
2-4 and 4-10 keV bands being $\gtsimm 1''$ in extent (see
Fig.~\ref{fig:cavrun28_bbimage}).

Comparing models R4 and R9 gives insight into the effects of reducing
the stellar wind speed, $v_{\rm S}$, with all other parameters kept
the same. In a similar respect to models R1 and R8, the reduction in
$v_{\rm S}$ causes the reverse shock to be closer to the star, and the
postshock stellar wind temperature to be lower
($\sim10^{6.5}\;$K). Column densities are slightly higher for model
R9. The $L_{\rm int}$'s are very similar, yet the $L_{\rm att}$'s and
$\eta$ are slightly lower. Comparison of the spectra shows that for
model R9 the flux level is slightly lower and the spectral slope at $E
\gtsimm 4\;$keV is slightly softer than model R4.

A number of known objects possessing outflow cavities where X-ray
emission has been detected have been inferred as early B-type
stars. These stars will have lower terminal wind speeds and mass-loss
rates than the O-type star used as our fiducial model. In model R10 we
use stellar wind parameters corresponding to a B1V star
\citep{Prinja:1989}. Consequently, the mass infall rate must be scaled
down to produce a similar ram pressure balance to that of model
R1. The evolution of the cavity is qualitatively similar to model
R1. However, the different ram pressures of the inflow and outflow
produces more vigourous ablation of the base of the cavity wall than
in model R1. The reduced stellar wind velocity results in a lower
maximum postshock gas temperature, which in-turn reduces the amount of
hard X-ray emission. The column densities are significantly reduced
compared to model R1, and are in agreement with observationally
determined values for embedded early B-type stars
\citep[e.g.][]{Preibisch:2002}. The spectrum is much softer now
(Fig.~\ref{fig:spectra}).

\vspace{-3mm}
\subsection{Variation with wind geometry}

Varying the wind geometry affects the contributions to the total
emission from the different components, and changes the cavity
evolution. For instance, increasing the opening angle of the stellar
wind provides more stellar wind material, and less disk wind
material. If the ram pressure of the stellar wind is much greater than
that from the disk wind the latter may be channelled into an
equatorial outflow, as found in model R8. To explore the consequences
of different wind geometries we have calculated models R11, R12, and
R13 (see Table~\ref{tab:models}).

As the opening angle of the stellar wind is increased (c.f. models R1,
R11, and R12), and that of the disk wind decreased, the thickness of
the disk wind layer along the cavity wall becomes thinner due to the
reduced amount of disk wind being injected. The evolution of the
cavity remains similar to that of R1, although the cavity wall is now
more curved towards the base. The base of the cavity wall is closer to
the star because the ram pressure of the outflow close to the disk
plane is now reduced. The aspect ratio of the region of unshocked
winds increases as the reverse shock moves to smaller $r$. The value
of $\eta$ does not vary much from that calculated for R1 though the
observed spectrum is harder (see the spectrum from model R12 in
Fig.~\ref{fig:spectra}).

To explore the effect of having a spherical outflow we have
constructed model R13, which is equivalent to R1 except with no disk
wind. We find that compared to model R1 there is a definite hardening
with more X-ray emission at $E\gtsimm4\;$keV (Fig.~\ref{fig:spectra}),
which is expected as there is now more high speed material in the
outflow. There is also an enhancement in $L_{\rm att_{C}}$ due to
heating of this gas to hotter X-ray emitting temperatures by direct
contact with, and thus greater turbulent heating from, the stellar
wind. The value of $\eta$ remains in agreement with observationally
detected values for MYSOs. Interestingly, this shows that X-ray
emission can be generated from the wind-cavity interaction in the
absence of a disk wind, which means this mechanism could be applicable
to more evolved protostars as well.

\vspace{-3mm}
\subsection{Variation with cavity opening angle}

The angle that the outflow makes with the cavity wall affects the rate
at which kinetic energy in the preshock gas is transferred into
thermal energy in the shocked gas. Also, the pressure of the shocked
gas is affected by the degree of confinement provided by the cavity
wall. Keeping all other parameters the same, a narrow cavity leads to
higher gas pressures than a wider cavity. This has implications for
the wind-driven expansion of the cavity and the position and shape of
the reverse shock. To explore these effects further models R14-R17
have been performed.

When $\omega$ is reduced relative to R1 (as in models R14 and R15) the
cavity is initially narrower, and the postshock thermal pressure
higher. This results in a rapid expansion of the cavity in order to
balance the pressure across the cavity wall. In model R14 the cavity
seems to have relaxed by $t=1000\;$yrs, whereas in model R15 expansion
still appears to be ongoing at this time. By the end of both
simulations the cavity has a roughly curved shape. The rapid expansion
of the cavity wall away from the star causes turbulent mixing and
fluctuations in the shape and position of the reverse
shock. Instabilities form in the cavity wall as the less dense outflow
collides with the more dense inflow, causing a combination of KH,
Rayleigh-Taylor, and Richtmeyer-Meshkov instabilities. The column
densities relative to those from model R1 are higher because of the
longer path lengths for lines-of-sight through cloud material and
because the initial cavity is cut into the cloud closer to the
star. The conclusion that the values of $\eta$ and $L_{\rm att_{T}}$
are higher for smaller initial cavity opening angles appears to only
be true for model R14. However, the lower $\eta$ at $t=1000\;$yr
resides at a minimum in the lightcurve for model R15. The spectrum
remains similar to that of R1 at $E\ltsimm 2\;$keV, and shows more
emission above this energy (Fig.~\ref{fig:spectra}).

\begin{figure}
\begin{center}
    \begin{tabular}{l}
      \resizebox{60mm}{!}{{\includegraphics{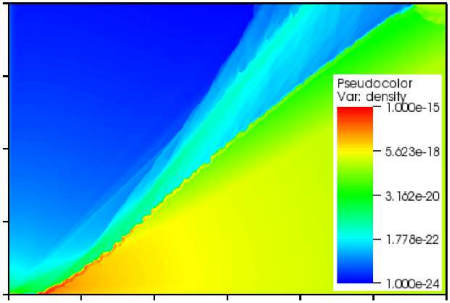}}}
      \\
    \end{tabular}
    \caption[]{Density snapshot taken from model R17 at
    $t=1000\;$yr. The tick marks on the axis correspond to a distance
    of $10^{16}\;$cm.}
    \label{fig:noreverseshock}
\end{center}
\end{figure}

In models R16 and R17 the cavity half opening angle is increased
compared to model R1. For $\omega=45^{\circ}$ (R16), a reverse shock
is still generated close to the star which encloses the unshocked
winds. Interestingly, when $\omega=60^{\circ}$ (R17) this no longer
happens (see Fig.~\ref{fig:noreverseshock}). This is the result of a
shallow grazing angle for the wind-cavity collision. The shocked gas
now only resides along the cavity wall, and the maximum temperatures
reached are $T\sim10^{6.5}\;$K. The disk wind forms a layer along the
cavity wall. On the interior side of the cavity it is bordered by the
shocked stellar wind. Column densities and the intrinsic and
attenuated emission decrease with increasing $\omega$. When there is
no reverse shock (close to the star) $\eta$ drops significantly and
the spectrum becomes much softer (see the spectrum from model R17 in
Fig.~\ref{fig:spectra}).

\vspace{-3mm}
\subsection{Variation with cavity curve}

\begin{figure}
\begin{center}
    \begin{tabular}{ll}
      \resizebox{40mm}{!}{{\includegraphics{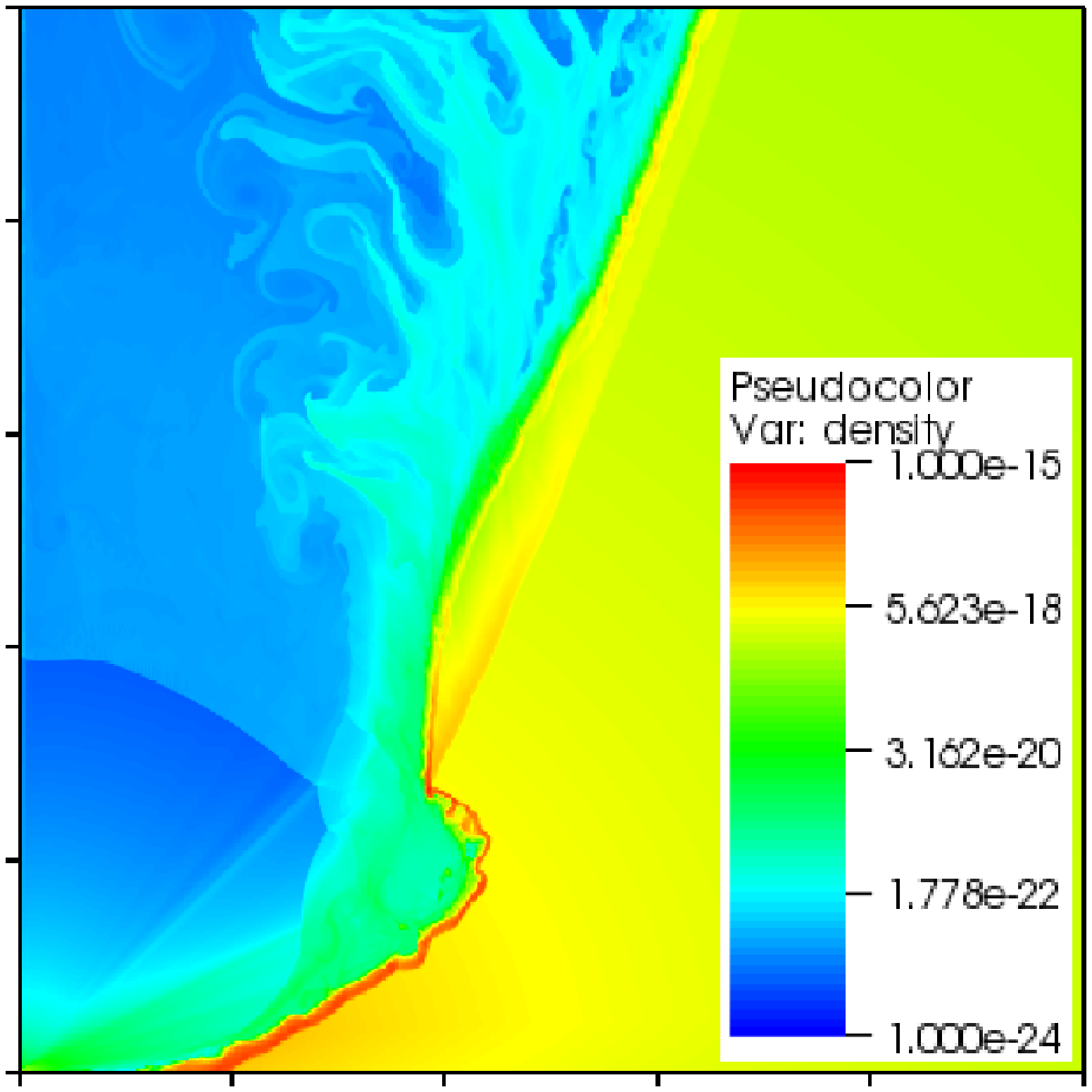}}} &
      \resizebox{40mm}{!}{{\includegraphics{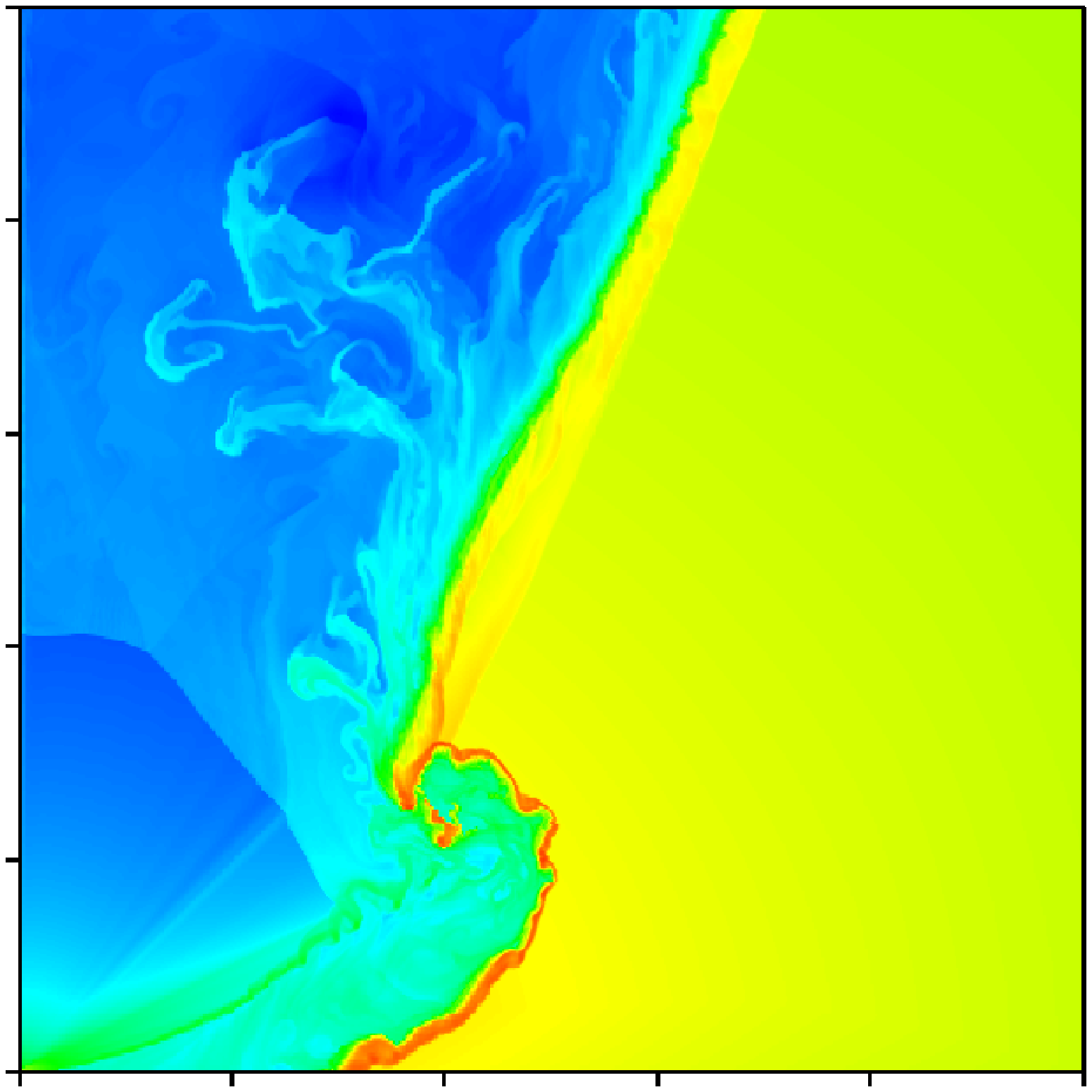}}} \\
    \end{tabular}
    \caption[]{Density snapshots taken from model R18 at $t=1400$
    (left) and 2000 yrs (right). The tick marks on the axis correspond
    to a distance of $10^{16}\;$cm.}
    \label{fig:curvehole}
\end{center}
\end{figure}

The shape of the cavity may not initially be conical, as has been
assumed in simulations R1-R17 and R19-20. In fact, some observations
of cavities around MYSOs show them to be more parabolic in shape. It
is unclear whether an initially conical outflow cavity evolves to
become more curved. Such evolution could be driven by the action of
the outflowing winds (as in R1), or by changes to the mechanism that
drives the outflow causing the resulting cavity to be curved at the
base. The initial curvature of the cavity wall is controlled by the
exponent $\beta$ in Eq.~\ref{eqn:cavity}. In model R18 we adopted
$\beta = 2$, i.e. a parabolic cavity. A reverse shock is still formed,
although it occurs at a greater distance from the star, and with a
different shape than that seen in model R1
(Fig.~\ref{fig:curvehole}). The reverse shock can be divided into two
parts. The upper part (which spans polar angles $0<\theta\ltsimm
60^{\circ}$) is very clearly normal to the stellar wind, which leads
to postshock gas at very high ($\sim10^{8}\;$K) temperatures and hard
and luminous X-ray emission (Fig.~\ref{fig:spectra}). The lower part,
in contrast, is very oblique to the disk wind, and leads to much
cooler postshock temperatures. For the duration of the simulation
infalling cloud material can still reach the disk plane. However, by
$t=2000\;$yr this accretion flow has halted due to the pocket of hot
rarefied disk wind gas which sweeps up the cloud material and drives
it into the equatorial outflow cavity (Fig.~\ref{fig:curvehole}). Also
noticeable by the end of the simulation is that the disk wind is
confined to the equatorial outflow bubble, and the larger cavity
towards the pole is filled with turbulent shocked stellar wind gas.

In conclusion, introducing curvature into the cavity wall at the start
of the simulation does not prevent integrated count rates in
approximate agreement with observations from being produced and the
spectrum appears largely similar, though a factor of 3 or so brighter
(Fig.~\ref{fig:spectra}). Curvature produced by a previous outflow may
be erased as the winds carve into the cavity wall.

\vspace{-3mm}
\subsection{Variation with centrifugal radius}

The spherically averaged density goes as $\rho \propto r^{-1/4}$
within $r_{\rm c}$, and $\rho \propto r^{-3/2}$ at distances greater
than $r_{\rm c}$. Therefore, changing $r_{\rm c}$ affects the density
distribution in the cloud.

In model R19, $r_{\rm c}$ is decreased to $0.8\times10^{15}\;$cm, and
as such the density in the cloud material should be lower than that of
model R1 at equivalent distances from the star. There is however
little noticeable difference between the cavity evolution in this
simulation and R1. There is a minor increase in $N_{\rm H-star}$ which
is consistent with the cavity wall being slightly closer to the star
for model R19. The increase in $N_{\rm EWC}$ is a result of the
spatial distribution of the emission and the position of the reverse
shock. The reverse shock is now closer to the star than in model R1,
and as such there is more emission from the disk wind close to the
base of the cavity wall. Lines-of-sight to this disk wind gas pass
through the dense cloud gas close to the axis, resulting in higher
column densities. However, in spite of this $\eta$ is higher because
of the enhancement to the emission measure brought about by the
position of the reverse shock.

The value of $r_{\rm c}$ is increased to $6.0\times10^{15}\;$cm in
simulation R20. The winds now excavate the base of the cavity wall
more. This excavation proceeds in an unsteady manner, and results in
more variation in the degree of mixing between the winds and the
position of the reverse shock throughout the simulation than seen in
model R1. These differences in the dynamics result in increased
variability of the resulting emission; for model R1 the largest
amplitude oscillations in the attenuated luminosities was a factor of
$\sim 2$ (Fig.~\ref{fig:lcurves}), whereas for model R20 there are
oscillations of a factor of $\sim 4$. The spectrum at $t=1000\;$yrs
represents the median for model R20. 

\vspace{-3mm}
\section{Candidate objects}
\label{sec:fitstoobjects}

We compare our model against X-ray observations of three objects: S106
IRS4, Mon R2 IRS3 A, and AFGL 2591. Because of the low count rates
spectral fits to these objects are relatively meaningless. Therefore
we restrict our comparison to the integrated count rates and the
generation of X-rays above 2 keV. We also examine whether the model
flux falls below the detection limit (and is spatially unresolvable)
when reasonable model parameters are used. We place the constraint on
the model parameters that the cavity must be reasonably steady
(i.e. the infall does not constrict the outflow at any point during
the simulation) and the visual extinction agrees with
observation. Therefore, the main free parameters are $\dot{M}_{\rm S}$
and $\dot{M}_{\rm infall}$ (as disk wind parameters were based on
previous observational estimates).

\begin{table}
\begin{center}
\caption[]{Model parameters (see also Table~\ref{tab:models}) and
results for the candidate objects. The distances are taken from
\cite{Schneider:2006}, \cite{Staude:1982} and \cite{Racine:1968}, and
the inclination angles are from \cite{vanderTak:2006},
\cite{Solf:1982} and \cite{alvarez:2004a} for AFGL 2591, S106 IRS4, and
Mon R2 IRS3 A, respectively. $N_{\rm H-star}$ and $N_{\rm EWC}$ are in
units of $10^{23}\;{\rm cm^{-2}}$.}
\begin{tabular}{llllll}
\hline
Model & D & $i$ & $N_{\rm H-star}$ & $N_{\rm EWC}$ & $\eta$ \\
 & (kpc) & ($^{\circ}$) &  &  & (ks$^{-1}$)\\
\hline
S106 & 0.6 & 75 & 0.4 & 0.8 & 0.23 \\
Mon R2 & 0.83 & 45 & 0.6 & 1.1 & 0.12 \\
AFGL 2591 & 1.7 & 32 & 2.0 & 6.9 & 0.09 \\
\hline
\label{tab:candidates}
\end{tabular}
\end{center}
\end{table}

\vspace{-3mm}
\subsection{S106 IRS4}

S106, at a distance of 600$\pm$100 pc \citep{Staude:1982}, is a
massive star-forming region known for an extended bipolar HII region,
which is illuminated by the $\gtsimm 15 \Msol$ massive star S106 IRS4
\citep{Felli:1984}. \cite{Schneider:2002} find the observed morphology
and kinematics in $^{12}$CO and $^{13}$CO 2$\rightarrow$1 to be
consistent with a cavity created by the radiation and ionized wind
from S106 IRS4 sweeping up material from the cavity wall, marked by
the two lobes of the HII region and at the extreme ends of the
flow. They also find that the double-peak structure of the cloud
breaks down and the emission merges into a more diffuse extended
plateau at S106 IRS4, where the molecular emission traces the
red-shifted component of the stellar wind hitting the backside of the
cavity walls. Near-IR speckle observations by \cite{alvarez:2004b}
show only a single unresolved source at the position of S106
IRS4. \cite{Solf:1982} used the kinematics of the bipolar nebula to
deduce that the inclination angle of the system in the plane of the
sky is $\sim75^{\circ}$. 

To determine the input parameters for our model of S106 we calculated
a series of models with various $\dot{M}_{\rm S}$, $\dot{M}_{\rm D}$,
and $\dot{M}_{\rm infall}$, where we fixed $i$, $v_{\rm S}$, $v_{\rm
D}$, and $\omega$. A good match to the \textit{Chandra} count rate and
the visual extinction, $A_{\rm v}$, was found for one particular
model, with parameters as noted in Table~\ref{tab:models}. From this
model we calculate a count rate (Table~\ref{tab:candidates}) in
agreement with the 0.30$\pm$0.11 ks$^{-1}$ \textit{Chandra} detection
\citep{Giardino:2004}. The $N_{\rm H-star}$ and $N_{\rm EWC}$ values
correspond to visual extinctions of $A_{\rm v}=22$ and 39 mag
respectively, which are consistent with previously determined values
\citep{Giardino:2004}. The model parameters for the stellar wind
terminal velocity and mass-loss rate are consistent with an early
B-type star. For the disk wind, the terminal velocity was set to the
previous estimate of $\simeq350\;{\rm km\thinspace s}^{-1}$
\citep{Drew:1993}, whereas the mass-loss rate is roughly an order of
magnitude lower than previously determined values of
$1.6-1.8\times10^{-6}\Msolpyr$
\citep{Felli:1984,Hoare:1994,Gibb:2007}. This difference may be due to
wind clumping causing over-estimates in the mass-loss rate
determinations from free-free emission. We note that amongst B-type
stars, a mass-loss rate as high as $1.8\times10^{-6}\Msolpyr$ is
consistent only with an evolved supergiant luminosity type, with main
sequence stars having mass-loss rates typically two or more orders of
magnitude lower \citep[e.g.][]{Prinja:1989}.

\vspace{-5mm}
\subsection{Mon R2 IRS3 A}

The general structure of the observable nebula around IRS3~A is
clearly monopolar, which suggests IRS3 A is embedded in a disk or
torus with bipolar cavities. \cite{Preibisch:2002} obtained high
resolution (75 mas = 62 au) near-infrared \textit{H} and \textit{K}
band images of Mon R2 IRS 3 which show a close triple system
surrounded by strong diffuse nebulosity and three additional infrared
sources within 3'' of the brightest object IRS3 A ($K\sim7.9$,
$M_{\ast} = 12-15 \Msol$), which has not yet developed an HII region
\citep{alvarez:2004b}. For a review of the Mon R2 star-forming region
see \cite{Carpenter:2008}. \cite{Preibisch:2002} comment that the
X-ray properties of IRS3 A and C cannot be explained by stellar wind
models where the X-rays are generated at shocks inherent to the winds
as the emission is much harder. This agrees with the comment by
\cite{Kohno:2002} that the observed plasma temperatures of $\gtsimm
4\;$keV ($T\gtsimm5\times10^{7}\;$K) are ten times higher than plasma
temperatures typical for early B-type main sequence stars. As has been
shown in \S~\ref{sec:results}, shock temperatures in a winds-cavity
interaction can reach $\gtsimm 10^{8}\;$K.

In model Mon R2 (Table~\ref{tab:models}) and the subsequent X-ray
calculation we use values for $r_{\rm c}$, $r_{\rm cav}$, $\omega$,
and $i$ determined by \cite{alvarez:2004a} from fits to observations
with envelope radiative transfer models. The model count rate
(Table~\ref{tab:candidates}) agrees with the \textit{Chandra} value of
0.166$\pm$0.041 ks$^{-1}$ determined by \cite{Preibisch:2002}. The
stellar wind terminal velocity and mass-loss rate imply a
late-O/early-B type central star. The mass infall rate for model Mon
R2 is roughly a factor of 5 lower than that determined by
\cite{alvarez:2004a}. However, this difference is not surprising as
the evolved state of the envelope in our model produces a dense layer
of compressed cloud material along the cavity wall, which increases
the column to the star. This feature is not present in envelope
radiative transfer models, which typically assume a smooth density
distribution. As such, to attain a visual extinction which agrees with
the observations we must decrease the mass-infall rate.

\vspace{-3mm}
\subsection{AFGL 2591}

The $\sim16\Msol$ \citep{vanderTak:2005} early B-type star AFGL 2591
sits at the centre of a powerful bipolar molecular outflow cavity
orientated in the east-west direction
\citep{Lada:1984,Mitchell:1992}. The central source is fully obscured
in the optical and \textit{J} band. The existence of a large-scale
torus perpendicular to the outflow cavity is inferred from extended CS
emission \citep{Yamashita:1987}. Nebulous loops are clearly seen in
the bispectrum speckle interferometry images of \cite{Preibisch:2003},
which are thought to mark the peripheries of outflow bubbles. When
modelling high-resolution SO data, \cite{Benz:2007} found the
inclusion of X-ray emission (which may originate from outflow shocks)
improved model fits. 

AFGL~2591 was observed by {\it Chandra} on 2006 February 8 using the
Advanced CCD Imaging Spectrometer (ACIS) S3 chip in very faint mode
for an exposure time of 30.17~ks. Data were obtained from the {\it
Chandra Data Archive}\footnote{http://cxc.harvard.edu/cda/} and were
processed in a standard way using the {\it Chandra} Interactive
Analysis of Observations \citep[\textsc{CIAO,}][]{Fruscione:2006}
software version 4.1.1.

\begin{figure}
\begin{center}
    \begin{tabular}{l}
      \resizebox{85mm}{!}{{\includegraphics{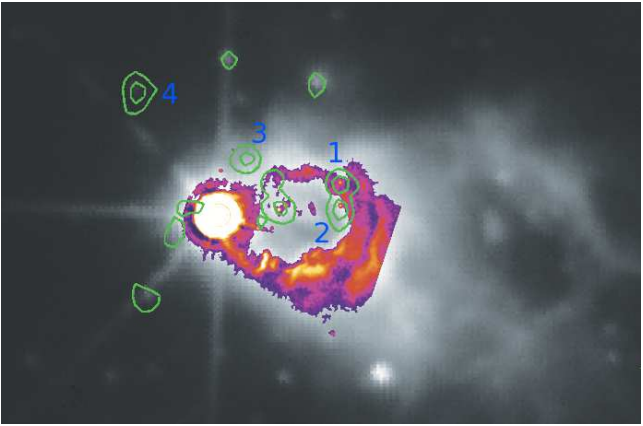}}} \\
    \end{tabular}
    \caption[]{Near-infrared speckle interferometry from
  \cite{Preibisch:2003} and contours from a {\it Chandra} X-ray image
  smoothed with a 1'' gaussian super-imposed on the UKIDSS \textit{K}
  band image of the AFGL 2591 region. The detected point-like sources
  noted in Table~\ref{tab:chandrapoint} are labelled.}
    \label{fig:chandra}
\end{center}
\end{figure}

Point source detection was performed using the CIAO {\tt wavdetect}
task \citep{Freeman:2002} on the level 2 event file. The process works
by correlating the dataset with a number of ``Mexican Hat'' wavelet
functions at different spatial scales to search for correlations. This
process was performed at a significance threshold of $10^{-5}$,
corresponding to less than 1 false source for the whole S3 field, and
resulted in 4 sources detected in the vicinity of AFGL 2591
(Fig.~\ref{fig:chandra} and Table~\ref{tab:chandrapoint}) and 62
sources in the whole S3 field. The point sources are identified with
stellar point sources in a deep UKIDSS \textit{K} band image of the
region illustrated in Fig.~\ref{fig:chandra}.

To correct for systematic errors in the Chandra aspect system
(typically about 1 arcsec), we searched for near-IR counterparts from
the UKIDSS \citep{Lawrence:2007} Galactic Plane Survey
\citep{Lucas:2008} catalogue across the entire CCD and identified
bright counterparts for 12 of the detected X-ray point sources.  These
counterparts were used to correct the {\it Chandra} coordinates and
register them to the UKIDSS reference frame.

Source extraction was performed using {\sc ACIS Extract}
\citep[AE,][]{Broos:2002}, an IDL-based package developed for ACIS
data processing. The procedures used in AE are described in
\citet{TFM:03} and \citet{Getman:05}. Model point spread functions
(PSFs) were constructed for each source together with contours which
contain $>$90\%\ of source events. The background was estimated
locally for each source using circular annuli around their respective
PSFs that avoided the PSFs of other nearby sources. From the
background subtracted net counts, the median energy, source
significance and the Poisson ``not-a-source'' probability were
calculated for each source. The results, source positions, and
characteristics are listed in Table~\ref{tab:chandrapoint}. These
sources are neighbouring low-mass stars and are not associated with
the outflow from AFGL 2591.

\begin{table}
 \center
  \caption{Detected point-like X-ray sources in the vicinity of
  AFGL2591. $\tilde{E}$ is the median photon energy from the
  source. The net counts have been background subtracted. $P_{\rm
  not}$ is the Poisson probability that the source is a chance
  coincidence of background photons---the ``not-a-source''
  probability.}
\label{tab:chandrapoint}
\begin{tabular}{lllllll}
\hline
No.	& RA	 & Dec & $\tilde{E}$ & Net & Signif.
  & $P_{\rm not}$ \\
   & (deg) & (deg) & (keV) & counts &  ($\sigma$) &  \\
\hline
1 & 307.3508 & 40.1894  & 3.9 &	6.69 &	1.77 & 	5.7e-8 \\
2 & 307.3508 & 40.1888  & 3.2 &	6.71 &	1.77 &  3.3e-8 \\
3 & 307.3554 & 40.1910  & 2.9 &	6.72 &	1.78 & 2.6e-8 \\
4 & 307.3530 & 40.1898  & 4.3 &	5.72 &	1.59 & 6.6e-7 \\
\hline
\end{tabular}
\end{table}

AFGL 2591 was not detected as a point source, unlike S106 IRS4 and Mon
R2 IRS3 A where the X-ray emission was unresolved. However, there is a
tentative detection of diffuse emission (visible as the contours east
of the central source in Fig.~\ref{fig:chandra}) with count rate of
$\sim0.2\;$ks$^{-1}$ from a circular region, centered on AFGL 2591
with radius $\sim2.6''$, with a significance of 3.7~$\sigma$ above the
background level \citep[determined using the method of][which
expresses the probability that the observed counts cannot be explained
by Poissonian variations in the background and is different to the
signal-to-noise ratio of the extracted signal]{Pease:2006}. The origin
of the observed emission is unclear and we interpret it as due to the
interaction of the winds from the central object with the surrounding
envelope. 

In our model of AFGL 2591 (Table~\ref{tab:models}) we use a mass
infall rate determined from a comparison of Eq.~\ref{eqn:density} with
the \cite{vanderTak:1999} expression for number density and assuming
the infalling cloud material to be molecular hydrogen. We adopt a
value of $15^{\circ}$ for the cavity half opening angle
\citep[see][]{vanderTak:1999,Poelman:2007}. The stellar wind velocity
is chosen to agree with an early B-type star. The disk wind velocity
agrees well with the $\sim500\;{\rm km
\thinspace s^{-1}}$ velocities inferred from the observed
high-velocity wings of infrared absorption lines
\citep{vanderTak:1999}. Our model results show a count rate of
$0.09\;$ks$^{-1}$ which is consistent with the conservative
upper-limit of $0.2\;$ks$^{-1}$. The visual exctinction to the star
($A_{\rm v}\simeq105\;$mag) agrees well with the 100 mag determined by
\cite{vanderTak:1999} from JCMT observations of the C$^{17}$O
J=2$\rightarrow$1 and J=3$\rightarrow$2 lines. The molecular gas with
velocities of $\sim 200\;{\rm km\thinspace s^{-1}}$ observed by
\cite{vanderTak:1999} and interpreted as the entrainment of cloud
material by the ionized wind outflow can be readily explained by our
winds-cavity model.

\vspace{-3mm}
\section{Discussion}
\label{sec:discussion}

For model R1 the efficiencies for converting the mechanical wind power
into thermal energy and then into X-ray emission ($=L_{\rm int} /
\zeta$) are $\simeq 5\times10^{-6}$ and $\simeq 9\times10^{-4}$ for
the stellar and disk winds respectively, where $\zeta$ is the power of
the specified wind. These values are representative of those
calculated for all of the models with a common trend being that the
efficiency is roughly two orders of magnitude higher for the disk
wind. In many cases the disk wind dominates the attenuated X-ray
emission, despite the mechanical power being typically less than that
from the faster stellar wind. However, the faster stellar wind is
always responsible for the hardest X-ray emission.

The model can provide a match to the observed X-ray characteristics of
a sample of MYSOs. The intrinsic variability to the observed emission,
caused by the dynamics and fluctuations in the position of the reverse
shock, complicates placing constraints on model parameters. One
possible avenue for future work would be to attempt simultaneous fits
to radio observations with the models.

A recurring feature of the simulations was the delicate balance
between the infall and the outflow. It proved impossible to attain a
set of parameters where the cavity wall remained stable and roughly
stationary for $t\simeq10^{4}\;$yr. The cause of this problem is the
imbalance between the pressure in the infalling material and that in
the outflows. Noting that the method used to describe the shape of the
cavity wall is essentially arbitrary, a more realistic approach would
be to have an initially fully embedded MYSO and to create a
self-consistent cavity via a jet/outflow.

Intuitively one would expect the opening angle of a cavity to be tied
to the evolution of the central star(s), and that an initially narrow
cavity resulting from a molecular outflow would be widened by outflows
as the star evolves to finally be revealed as a main sequence star
\citep[e.g.][]{Velusamy:1998}. In a recent paper, \cite{Canto:2008}
modelled outflow cavities using a prescription with a time dependent
opening-angle for the outflow. Although observational results could be
qualitatively reproduced, it is unclear from the wind-cavity models
whether such an approach is physically realistic.

The model used in this paper considers the central object to be a
single star. However, most massive stars form in binaries, which
allows the collision of winds between the stars as well
\citep[e.g.][]{Stevens:1992,Parkin:2008, Pittard:2009}, and could
potentially generate considerably more X-ray emission
\citep[e.g.][]{Parkin:2009}.

\vspace{-3mm}
\section{Conclusions}
\label{sec:conclusions}

Hydrodynamical simulations of the wind-cavity interaction around an
embedded MYSO have been studied. The latitude dependent wind geometry
we use incorporates both a stellar and disk wind. Using an extensive
range of simulations we have examined the effect of varying different
model parameters on the evolution of the cavity and the resulting
observational characteristics (i.e. column densities, X-ray
luminosities and count rates). The main conclusions from this work
are:

\begin{itemize}
\item The collision of the winds against the cavity wall generates a
  reverse shock (for cavity half opening angles $\omega\ltsimm
  60^{\circ}$) close to the star ($\ltsimm 500\;$au). The shock heated
  gas produces X-ray emission with an integrated count rate and
  spatial extent in agreement with observations of MYSOs by
  \textit{Chandra}. The position and shape of the reverse shock is
  dependent on the ram pressure in the inflow/outflow. Fluctuations in
  the position of the reverse shock cause variability of the observed
  emission on timescales of several hundred years, and possibly on
  shorter timescales which were not probed.
\item The amount of X-ray emission in the 4-10 keV band is dependent
  on the position of the reverse shock, which is strongly related to
  the stellar wind speed and the adopted mass inflow and outflow
  rates.
\item Integrated count rates in agreement with \textit{Chandra}
  detections of MYSOs are obtained across a wide region of model
  parameter space, indicating that the generation of X-ray emission
  through the interaction of an outflow with infalling material is
  potentially a very robust process.
\item There is a limiting opening angle of the cavity for which a
  reverse shock is produced which resides close to the star and
  encloses the unshocked winds. For our adopted wind geometry we find
  this (full) opening angle to be $120^{\circ}$.
\item There appears to be a delicate hydrodynamic balance between the
inflow and outflow.
\end{itemize}

The model presented in this work provides a useful first insight into
the interaction between winds from an MYSO and the surrounding
envelope. As with many astrophysical problems, there is a vast range
of scales to consider in modelling the wind-cavity interaction, and
the methods employed dictate the approximations that must be made. To
accurately model the driving of the winds requires high resolution in
the wind acceleration region \citep[e.g.][]{Proga:1998}. Future models
would benefit greatly from higher resolution in the region close to
the star and disk. This would allow the interaction between the
outflow and inflow from the acceleration of the disk wind to the
intersection point between the disk wind and the cavity to be
examined. This would be an improvement on the current work as it would
allow the stellar and disk winds to be self-consistently driven and
the photoevaporation of the accretion disk
\citep[e.g.][]{Hollenbach:1994,Richling:2000} could be examined. On
larger scales, the evolution of the HII region around the star could
be followed.

\subsection*{Acknowledgements}
ERP thanks the University of Leeds for funding through a Henry Ellison
Scholarship. JMP gratefully acknowledges funding from the Royal
Society. We thank the anonymous referee for helpful comments which
improved the presentation of this paper.
\vspace{-3mm}

\begin{thebibliography}{}

\bibitem[\protect\citeauthoryear{{Alvarez}, {Hoare} \& {Lucas}}{{Alvarez}
  et~al.}{2004a}]{alvarez:2004a}
{Alvarez}, C., {Hoare}, M., \& {Lucas}, P. 2004a, \aap, 419, 203

\bibitem[\protect\citeauthoryear{{Alvarez}, {Hoare}, {Glindemann} \&
  {Richichi}}{{Alvarez} et~al.}{2004b}]{alvarez:2004b}
{Alvarez}, C., {Hoare}, M., {Glindemann}, A., \& {Richichi}, A. 2004b, \aap,
  427, 505

\bibitem[\protect\citeauthoryear{{Banerjee} \& {Pudritz}}{{Banerjee} \&
  {Pudritz}}{2006}]{Banerjee+Pudritz:2006}
{Banerjee}, R. \& {Pudritz}, R.~E. 2006, \apj, 641, 949

\bibitem[\protect\citeauthoryear{{Banerjee} \& {Pudritz}}{{Banerjee} \&
  {Pudritz}}{2007}]{Banerjee:2007}
{Banerjee}, R. \& {Pudritz}, R.~E. 2007, \apj, 660, 479

\bibitem[\protect\citeauthoryear{{Beltr{\'a}n}, {Cesaroni}, {Codella}, {Testi},
  {Furuya} \& {Olmi}}{{Beltr{\'a}n} et~al.}{2006}]{Beltran:2006}
{Beltr{\'a}n}, M.~T., {Cesaroni}, R., {Codella}, C., {Testi}, L., {Furuya},
  R.~S., \& {Olmi}, L. 2006, \nat, 443, 427

\bibitem[\protect\citeauthoryear{{Benz}, {St{\"a}uber}, {Bourke}, {van der
  Tak}, {van Dishoeck} \& {J{\o}rgensen}}{{Benz} et~al.}{2007}]{Benz:2007}
{Benz}, A.~O., {St{\"a}uber}, P., {Bourke}, T.~L., {van der Tak}, F.~F.~S.,
  {van Dishoeck}, E.~F., \& {J{\o}rgensen}, J.~K. 2007, \aap, 475, 549

\bibitem[\protect\citeauthoryear{{Beuther}, {Schilke}, {Gueth}, {McCaughrean},
  {Andersen}, {Sridharan} \& {Menten}}{{Beuther} et~al.}{2002}]{Beuther:2002b}
{Beuther}, H., {Schilke}, P., {Gueth}, F., {McCaughrean}, M., {Andersen}, M.,
  {Sridharan}, T.~K., \& {Menten}, K.~M. 2002, \aap, 387, 931

\bibitem[\protect\citeauthoryear{{Blondin}, {Kallman}, {Fryxell} \&
  {Taam}}{{Blondin} et~al.}{1990}]{Blondin:1990}
{Blondin}, J.~M., {Kallman}, T.~R., {Fryxell}, B.~A., \& {Taam}, R.~E. 1990,
  \apj, 356, 591

\bibitem[\protect\citeauthoryear{{Broos}, {Townsley}, {Getman} \&
  {Bauer}}{{Broos} et~al.}{2002}]{Broos:2002}
{Broos}, P., {Townsley}, L., {Getman}, K., \& {Bauer}, F. 2002, {ACIS Extract,
  An ACIS Point Source Extraction Package}.
Pennsylvania State University

\bibitem[\protect\citeauthoryear{{Broos}, {Feigelson}, {Townsley}, {Getman},
  {Wang}, {Garmire}, {Jiang} \& {Tsuboi}}{{Broos} et~al.}{2007}]{Broos:2007}
{Broos}, P.~S., {Feigelson}, E.~D., {Townsley}, L.~K., {Getman}, K.~V., {Wang},
  J., {Garmire}, G.~P., {Jiang}, Z., \& {Tsuboi}, Y. 2007, \apjs, 169, 353

\bibitem[\protect\citeauthoryear{{Bunn}, {Hoare} \& {Drew}}{{Bunn}
  et~al.}{1995}]{Bunn:1995}
{Bunn}, J.~C., {Hoare}, M.~G., \& {Drew}, J.~E. 1995, \mnras, 272, 346

\bibitem[\protect\citeauthoryear{{Cant{\'o}}, {Raga} \& {Williams}}{{Cant{\'o}}
  et~al.}{2008}]{Canto:2008}
{Cant{\'o}}, J., {Raga}, A.~C., \& {Williams}, D.~A. 2008, Revista Mexicana de
  Astronomia y Astrofisica, 44, 293

\bibitem[\protect\citeauthoryear{{Carpenter} \& {Hodapp}}{{Carpenter} \&
  {Hodapp}}{2008}]{Carpenter:2008}
{Carpenter}, J. \& {Hodapp}, K. 2008, ArXiv e-prints

\bibitem[\protect\citeauthoryear{{Cassen} \& {Moosman}}{{Cassen} \&
  {Moosman}}{1981}]{Cassen:1981}
{Cassen}, P. \& {Moosman}, A. 1981, Icarus, 48, 353

\bibitem[\protect\citeauthoryear{{Chevalier}}{{Chevalier}}{1983}]{Chevalier:19%
83}
{Chevalier}, R.~A. 1983, \apj, 268, 753

\bibitem[\protect\citeauthoryear{{Colella} \& {Woodward}}{{Colella} \&
  {Woodward}}{1984}]{Colella:1984}
{Colella}, P. \& {Woodward}, P.~R. 1984, Journal of Computational Physics, 54,
  174

\bibitem[\protect\citeauthoryear{{Cox}}{{Cox}}{2000}]{Cox:2000}
{Cox}, A.~N. 2000, {Allen's astrophysical quantities}.
Allen's Astrophysical Quantities

\bibitem[\protect\citeauthoryear{{Crapsi}, {van Dishoeck}, {Hogerheijde},
  {Pontoppidan} \& {Dullemond}}{{Crapsi} et~al.}{2008}]{Crapsi:2008}
{Crapsi}, A., {van Dishoeck}, E.~F., {Hogerheijde}, M.~R., {Pontoppidan},
  K.~M., \& {Dullemond}, C.~P. 2008, \aap, 486, 245

\bibitem[\protect\citeauthoryear{{Cunningham}, {Frank} \&
  {Hartmann}}{{Cunningham} et~al.}{2005}]{Cunningham:2005}
{Cunningham}, A., {Frank}, A., \& {Hartmann}, L. 2005, \apj, 631, 1010

\bibitem[\protect\citeauthoryear{{Davis}, {Varricatt}, {Todd} \& {Ramsay
  Howat}}{{Davis} et~al.}{2004}]{Davis:2004}
{Davis}, C.~J., {Varricatt}, W.~P., {Todd}, S.~P., \& {Ramsay Howat}, S.~K.
  2004, \aap, 425, 981

\bibitem[\protect\citeauthoryear{{Delamarter}, {Frank} \&
  {Hartmann}}{{Delamarter} et~al.}{2000}]{Delamarter:2000}
{Delamarter}, G., {Frank}, A., \& {Hartmann}, L. 2000, \apj, 530, 923

\bibitem[\protect\citeauthoryear{{Dougherty}, {Pittard}, {Kasian}, {Coker},
  {Williams} \& {Lloyd}}{{Dougherty} et~al.}{2003}]{Dougherty:2003}
{Dougherty}, S.~M., {Pittard}, J.~M., {Kasian}, L., {Coker}, R.~F., {Williams},
  P.~M., \& {Lloyd}, H.~M. 2003, \aap, 409, 217

\bibitem[\protect\citeauthoryear{{Drew}, {Bunn} \& {Hoare}}{{Drew}
  et~al.}{1993}]{Drew:1993}
{Drew}, J.~E., {Bunn}, J.~C., \& {Hoare}, M.~G. 1993, \mnras, 265, 12

\bibitem[\protect\citeauthoryear{{Drew}, {Proga} \& {Stone}}{{Drew}
  et~al.}{1998}]{Drew:1998}
{Drew}, J.~E., {Proga}, D., \& {Stone}, J.~M. 1998, \mnras, 296, L6+

\bibitem[\protect\citeauthoryear{{Felli}, {Massi}, {Staude}, {Reddmann},
  {Eiroa}, {Hefele}, {Neckel} \& {Panagia}}{{Felli} et~al.}{1984}]{Felli:1984}
{Felli}, M., {Massi}, M., {Staude}, H.~J., {Reddmann}, T., {Eiroa}, C.,
  {Hefele}, H., {Neckel}, T., \& {Panagia}, N. 1984, \aap, 135, 261

\bibitem[\protect\citeauthoryear{{Freeman}, {Kashyap}, {Rosner} \&
  {Lamb}}{{Freeman} et~al.}{2002}]{Freeman:2002}
{Freeman}, P.~E., {Kashyap}, V., {Rosner}, R., \& {Lamb}, D.~Q. 2002, \apjs,
  138, 185

\bibitem[\protect\citeauthoryear{{Fruscione}, {McDowell}, {Allen},
  {Brickhouse}, {Burke}, {Davis}, {Durham}, {Elvis}, {Galle} \&
  {Harris}}{{Fruscione} et~al.}{2006}]{Fruscione:2006}
{Fruscione}, A., {et~al.} 2006, in Society of Photo-Optical Instrumentation
  Engineers (SPIE) Conference Series Vol.~6270 of Society of Photo-Optical
  Instrumentation Engineers (SPIE) Conference Series, {CIAO: Chandra's data
  analysis system}

\bibitem[\protect\citeauthoryear{{Garay} \& {Lizano}}{{Garay} \&
  {Lizano}}{1999}]{Garay:1999}
{Garay}, G. \& {Lizano}, S. 1999, \pasp, 111, 1049

\bibitem[\protect\citeauthoryear{{Gardiner}, {Frank} \& {Hartmann}}{{Gardiner}
  et~al.}{2003}]{Gardiner:2003}
{Gardiner}, T.~A., {Frank}, A., \& {Hartmann}, L. 2003, \apj, 582, 269

\bibitem[\protect\citeauthoryear{{Giardino}, {Favata} \& {Micela}}{{Giardino}
  et~al.}{2004}]{Giardino:2004}
{Giardino}, G., {Favata}, F., \& {Micela}, G. 2004, \aap, 424, 965

\bibitem[\protect\citeauthoryear{{Gibb} \& {Hoare}}{{Gibb} \&
  {Hoare}}{2007}]{Gibb:2007}
{Gibb}, A.~G. \& {Hoare}, M.~G. 2007, \mnras, 380, 246

\bibitem[\protect\citeauthoryear{{Hoare}}{{Hoare}}{2006}]{Hoare:2006}
{Hoare}, M.~G. 2006, \apj, 649, 856

\bibitem[\protect\citeauthoryear{{Hoare}, {Drew}, {Muxlow} \& {Davis}}{{Hoare}
  et~al.}{1994}]{Hoare:1994}
{Hoare}, M.~G., {Drew}, J.~E., {Muxlow}, T.~B., \& {Davis}, R.~J. 1994, \apj,
  421, L51

\bibitem[\protect\citeauthoryear{{Hollenbach}, {Johnstone}, {Lizano} \&
  {Shu}}{{Hollenbach} et~al.}{1994}]{Hollenbach:1994}
{Hollenbach}, D., {Johnstone}, D., {Lizano}, S., \& {Shu}, F. 1994, \apj, 428,
  654

\bibitem[\protect\citeauthoryear{{Hosokawa} \& {Omukai}}{{Hosokawa} \&
  {Omukai}}{2009}]{Hosokawa:2009}
{Hosokawa}, T. \& {Omukai}, K. 2009, \apj, 691, 823

\bibitem[\protect\citeauthoryear{{Howarth} \& {Prinja}}{{Howarth} \&
  {Prinja}}{1989}]{Howarth:1989}
{Howarth}, I.~D. \& {Prinja}, R.~K. 1989, \apjs, 69, 527

\bibitem[\protect\citeauthoryear{{Kohno}, {Koyama} \& {Hamaguchi}}{{Kohno}
  et~al.}{2002}]{Kohno:2002}
{Kohno}, M., {Koyama}, K., \& {Hamaguchi}, K. 2002, \apj, 580, 626

\bibitem[\protect\citeauthoryear{{Krumholz}, {Klein}, {McKee}, {Offner} \&
  {Cunningham}}{{Krumholz} et~al.}{2009}]{Krumholz:2009}
{Krumholz}, M.~R., {Klein}, R.~I., {McKee}, C.~F., {Offner}, S.~S.~R., \&
  {Cunningham}, A.~J. 2009, ArXiv e-prints

\bibitem[\protect\citeauthoryear{{Krumholz}, {McKee} \& {Klein}}{{Krumholz}
  et~al.}{2005}]{Krumholz:2005}
{Krumholz}, M.~R., {McKee}, C.~F., \& {Klein}, R.~I. 2005, \apj, 618, L33

\bibitem[\protect\citeauthoryear{{Lada}, {Thronson} Jr., {Smith}, {Schwartz} \&
  {Glaccum}}{{Lada} et~al.}{1984}]{Lada:1984}
{Lada}, C.~J., {Thronson}, Jr., H.~A., {Smith}, H.~A., {Schwartz}, P.~R., \&
  {Glaccum}, W. 1984, \apj, 286, 302

\bibitem[\protect\citeauthoryear{{Lawrence}, {Warren}, {Almaini}, {Edge},
  {Hambly}, {Jameson}, {Lucas}, {Casali}, {Adamson} \& {Dye}}{{Lawrence}
  et~al.}{2007}]{Lawrence:2007}
{Lawrence}, A., {et~al.} 2007, \mnras, 379, 1599

\bibitem[\protect\citeauthoryear{{Lee}, {Stone}, {Ostriker} \& {Mundy}}{{Lee}
  et~al.}{2001}]{Lee:2001}
{Lee}, C.-F., {Stone}, J.~M., {Ostriker}, E.~C., \& {Mundy}, L.~G. 2001, \apj,
  557, 429

\bibitem[\protect\citeauthoryear{{Liedahl}, {Osterheld} \&
  {Goldstein}}{{Liedahl} et~al.}{1995}]{Liedahl:1995}
{Liedahl}, D.~A., {Osterheld}, A.~L., \& {Goldstein}, W.~H. 1995, \apj, 438,
  L115

\bibitem[\protect\citeauthoryear{{Lucas}, {Hoare}, {Longmore}, {Schr{\"o}der},
  {Davis}, {Adamson}, {Bandyopadhyay}, {de Grijs}, {Smith}, {Gosling} \&
  {Mitchison}}{{Lucas} et~al.}{2008}]{Lucas:2008}
{Lucas}, P.~W., {et~al.} 2008, \mnras, 391, 136

\bibitem[\protect\citeauthoryear{{Mendoza}, {Cant{\'o}} \& {Raga}}{{Mendoza}
  et~al.}{2004}]{Mendoza:2004}
{Mendoza}, S., {Cant{\'o}}, J., \& {Raga}, A.~C. 2004, Revista Mexicana de
  Astronomia y Astrofisica, 40, 147

\bibitem[\protect\citeauthoryear{{Mitchell}, {Hasegawa} \&
  {Schella}}{{Mitchell} et~al.}{1992}]{Mitchell:1992}
{Mitchell}, G.~F., {Hasegawa}, T.~I., \& {Schella}, J. 1992, \apj, 386, 604

\bibitem[\protect\citeauthoryear{{Owocki}, {Castor} \& {Rybicki}}{{Owocki}
  et~al.}{1988}]{Owocki:1988}
{Owocki}, S.~P., {Castor}, J.~I., \& {Rybicki}, G.~B. 1988, \apj, 335, 914

\bibitem[\protect\citeauthoryear{{Parkin} \& {Pittard}}{{Parkin} \&
  {Pittard}}{2008}]{Parkin:2008}
{Parkin}, E.~R. \& {Pittard}, J.~M. 2008, \mnras, 388, 1047

\bibitem[\protect\citeauthoryear{{Parkin}, {Pittard}, {Corcoran}, {Hamaguchi}
  \& {Stevens}}{{Parkin} et~al.}{2009}]{Parkin:2009}
{Parkin}, E.~R., {Pittard}, J.~M., {Corcoran}, M.~F., {Hamaguchi}, K., \&
  {Stevens}, I.~R. 2009, \mnras, 394, 1758

\bibitem[\protect\citeauthoryear{{Patel}, {Curiel}, {Sridharan}, {Zhang},
  {Hunter}, {Ho}, {Torrelles}, {Moran}, {G{\'o}mez} \& {Anglada}}{{Patel}
  et~al.}{2005}]{Patel:2005}
{Patel}, N.~A., {et~al.} 2005, \nat, 437, 109

\bibitem[\protect\citeauthoryear{{Pease}, {Drake} \& {Kashyap}}{{Pease}
  et~al.}{2006}]{Pease:2006}
{Pease}, D.~O., {Drake}, J.~J., \& {Kashyap}, V.~L. 2006, \apj, 636, 426

\bibitem[\protect\citeauthoryear{{Pittard}}{{Pittard}}{2007}]{Pittard:2007book}
{Pittard}, J.~M. 2007, {Mass-Loaded Flows}.
pp 245--+

\bibitem[\protect\citeauthoryear{{Pittard}}{{Pittard}}{2009}]{Pittard:2009}
{Pittard}, J.~M. 2009, \mnras, 396, 1743

\bibitem[\protect\citeauthoryear{{Poelman} \& {van der Tak}}{{Poelman} \& {van
  der Tak}}{2007}]{Poelman:2007}
{Poelman}, D.~R. \& {van der Tak}, F.~F.~S. 2007, \aap, 475, 949

\bibitem[\protect\citeauthoryear{{Preibisch}, {Balega}, {Schertl} \&
  {Weigelt}}{{Preibisch} et~al.}{2002}]{Preibisch:2002}
{Preibisch}, T., {Balega}, Y.~Y., {Schertl}, D., \& {Weigelt}, G. 2002, \aap,
  392, 945

\bibitem[\protect\citeauthoryear{{Preibisch}, {Balega}, {Schertl} \&
  {Weigelt}}{{Preibisch} et~al.}{2003}]{Preibisch:2003}
{Preibisch}, T., {Balega}, Y.~Y., {Schertl}, D., \& {Weigelt}, G. 2003, \aap,
  412, 735

\bibitem[\protect\citeauthoryear{{Prinja}}{{Prinja}}{1989}]{Prinja:1989}
{Prinja}, R.~K. 1989, \mnras, 241, 721

\bibitem[\protect\citeauthoryear{{Proga}}{{Proga}}{2003}]{Proga:2003}
{Proga}, D. 2003, \apj, 585, 406

\bibitem[\protect\citeauthoryear{{Proga}, {Stone} \& {Drew}}{{Proga}
  et~al.}{1998}]{Proga:1998}
{Proga}, D., {Stone}, J.~M., \& {Drew}, J.~E. 1998, \mnras, 295, 595

\bibitem[\protect\citeauthoryear{{Proga}, {Stone} \& {Drew}}{{Proga}
  et~al.}{1999}]{Proga:1999}
{Proga}, D., {Stone}, J.~M., \& {Drew}, J.~E. 1999, \mnras, 310, 476

\bibitem[\protect\citeauthoryear{{Racine}}{{Racine}}{1968}]{Racine:1968}
{Racine}, R. 1968, \aj, 73, 233

\bibitem[\protect\citeauthoryear{{Reipurth} \& {Bally}}{{Reipurth} \&
  {Bally}}{2001}]{Reipurth:2001}
{Reipurth}, B. \& {Bally}, J. 2001, \araa, 39, 403

\bibitem[\protect\citeauthoryear{{Richling} \& {Yorke}}{{Richling} \&
  {Yorke}}{2000}]{Richling:2000}
{Richling}, S. \& {Yorke}, H.~W. 2000, \apj, 539, 258

\bibitem[\protect\citeauthoryear{{Saxton}, {Bicknell}, {Sutherland} \&
  {Midgley}}{{Saxton} et~al.}{2005}]{Saxton:2005}
{Saxton}, C.~J., {Bicknell}, G.~V., {Sutherland}, R.~S., \& {Midgley}, S. 2005,
  \mnras, 359, 781

\bibitem[\protect\citeauthoryear{{Schneider}, {Bontemps}, {Simon}, {Jakob},
  {Motte}, {Miller}, {Kramer} \& {Stutzki}}{{Schneider}
  et~al.}{2006}]{Schneider:2006}
{Schneider}, N., {Bontemps}, S., {Simon}, R., {Jakob}, H., {Motte}, F.,
  {Miller}, M., {Kramer}, C., \& {Stutzki}, J. 2006, \aap, 458, 855

\bibitem[\protect\citeauthoryear{{Schneider}, {Simon}, {Kramer}, {Stutzki} \&
  {Bontemps}}{{Schneider} et~al.}{2002}]{Schneider:2002}
{Schneider}, N., {Simon}, R., {Kramer}, C., {Stutzki}, J., \& {Bontemps}, S.
  2002, \aap, 384, 225

\bibitem[\protect\citeauthoryear{{Shang}, {Allen}, {Li}, {Liu}, {Chou} \&
  {Anderson}}{{Shang} et~al.}{2006}]{Shang:2006}
{Shang}, H., {Allen}, A., {Li}, Z.-Y., {Liu}, C.-F., {Chou}, M.-Y., \&
  {Anderson}, J. 2006, \apj, 649, 845

\bibitem[\protect\citeauthoryear{{Shepherd}}{{Shepherd}}{2005}]{Shepherd:2005}
{Shepherd}, D. 2005, in {Cesaroni} R.,  {Felli} M.,  {Churchwell} E.,
  {Walmsley} M.,  eds, Massive Star Birth: A Crossroads of Astrophysics
  Vol.~227 of IAU Symposium, {Massive star outflows}.
pp 237--246

\bibitem[\protect\citeauthoryear{{Sim}, {Drew} \& {Long}}{{Sim}
  et~al.}{2005}]{Sim:2005}
{Sim}, S.~A., {Drew}, J.~E., \& {Long}, K.~S. 2005, \mnras, 363, 615

\bibitem[\protect\citeauthoryear{{Solf} \& {Carsenty}}{{Solf} \&
  {Carsenty}}{1982}]{Solf:1982}
{Solf}, J. \& {Carsenty}, U. 1982, \aap, 113, 142

\bibitem[\protect\citeauthoryear{{Staude}, {Lenzen}, {Dyck} \&
  {Schmidt}}{{Staude} et~al.}{1982}]{Staude:1982}
{Staude}, H.~J., {Lenzen}, R., {Dyck}, H.~M., \& {Schmidt}, G.~D. 1982, \apj,
  255, 95

\bibitem[\protect\citeauthoryear{{Stevens}, {Blondin} \& {Pollock}}{{Stevens}
  et~al.}{1992}]{Stevens:1992}
{Stevens}, I.~R., {Blondin}, J.~M., \& {Pollock}, A.~M.~T. 1992, \apj, 386, 265

\bibitem[\protect\citeauthoryear{{Strickland} \& {Blondin}}{{Strickland} \&
  {Blondin}}{1995}]{Strickland:1995}
{Strickland}, R. \& {Blondin}, J.~M. 1995, \apj, 449, 727

\bibitem[\protect\citeauthoryear{{Terebey}, {Shu} \& {Cassen}}{{Terebey}
  et~al.}{1984}]{Terebey:1984}
{Terebey}, S., {Shu}, F.~H., \& {Cassen}, P. 1984, \apj, 286, 529

\bibitem[\protect\citeauthoryear{{Torrelles}, {Patel}, {Curiel}, {Ho}, {Garay}
  \& {Rodr{\'{\i}}guez}}{{Torrelles} et~al.}{2007}]{Torrelles:2007}
{Torrelles}, J.~M., {Patel}, N.~A., {Curiel}, S., {Ho}, P.~T.~P., {Garay}, G.,
  \& {Rodr{\'{\i}}guez}, L.~F. 2007, \apj, 666, L37

\bibitem[\protect\citeauthoryear{{Ulrich}}{{Ulrich}}{1976}]{Ulrich:1976}
{Ulrich}, R.~K. 1976, \apj, 210, 377

\bibitem[\protect\citeauthoryear{{van der Tak} \& {Menten}}{{van der Tak} \&
  {Menten}}{2005}]{vanderTak:2005}
{van der Tak}, F.~F.~S. \& {Menten}, K.~M. 2005, \aap, 437, 947

\bibitem[\protect\citeauthoryear{{van der Tak}, {van Dishoeck}, {Evans} II,
  {Bakker} \& {Blake}}{{van der Tak} et~al.}{1999}]{vanderTak:1999}
{van der Tak}, F.~F.~S., {van Dishoeck}, E.~F., {Evans}, II, N.~J., {Bakker},
  E.~J., \& {Blake}, G.~A. 1999, \apj, 522, 991

\bibitem[\protect\citeauthoryear{{van der Tak}, {Walmsley}, {Herpin} \&
  {Ceccarelli}}{{van der Tak} et~al.}{2006}]{vanderTak:2006}
{van der Tak}, F.~F.~S., {Walmsley}, C.~M., {Herpin}, F., \& {Ceccarelli}, C.
  2006, \aap, 447, 1011

\bibitem[\protect\citeauthoryear{{Velusamy} \& {Langer}}{{Velusamy} \&
  {Langer}}{1998}]{Velusamy:1998}
{Velusamy}, T. \& {Langer}, W.~D. 1998, \nat, 392, 685

\bibitem[\protect\citeauthoryear{{Wang}, {Feigelson}, {Townsley},
  {Rom{\'a}n-Z{\'u}{\~n}iga}, {Lada} \& {Garmire}}{{Wang}
  et~al.}{2009}]{Wang:2009}
{Wang}, J., {Feigelson}, E.~D., {Townsley}, L.~K., {Rom{\'a}n-Z{\'u}{\~n}iga},
  C.~G., {Lada}, E., \& {Garmire}, G. 2009, \apj, 696, 47

\bibitem[\protect\citeauthoryear{{Wang}, {Townsley}, {Feigelson}, {Getman},
  {Broos}, {Garmire} \& {Tsujimoto}}{{Wang} et~al.}{2007}]{Wang:2007}
{Wang}, J., {Townsley}, L.~K., {Feigelson}, E.~D., {Getman}, K.~V., {Broos},
  P.~S., {Garmire}, G.~P., \& {Tsujimoto}, M. 2007, \apjs, 168, 100

\bibitem[\protect\citeauthoryear{{Wilkin} \& {Stahler}}{{Wilkin} \&
  {Stahler}}{2003}]{Wilkin:2003}
{Wilkin}, F.~P. \& {Stahler}, S.~W. 2003, \apj, 590, 917

\bibitem[\protect\citeauthoryear{{Wolfire} \& {Cassinelli}}{{Wolfire} \&
  {Cassinelli}}{1987}]{Wolfire:1987}
{Wolfire}, M.~G. \& {Cassinelli}, J.~P. 1987, \apj, 319, 850

\bibitem[\protect\citeauthoryear{{Yamashita}, {Sato}, {Tamura}, {Suzuki},
  {Kaifu}, {Takano}, {Mountain}, {Moore}, {Gatley} \& {Hough}}{{Yamashita}
  et~al.}{1987}]{Yamashita:1987}
{Yamashita}, T., {et~al.} 1987, \pasj, 39, 809

\bibitem[\protect\citeauthoryear{{Yorke} \& {Sonnhalter}}{{Yorke} \&
  {Sonnhalter}}{2002}]{Yorke:2002}
{Yorke}, H.~W. \& {Sonnhalter}, C. 2002, \apj, 569, 846

\bibitem[\protect\citeauthoryear{{Zinnecker} \& {Yorke}}{{Zinnecker} \&
  {Yorke}}{2007}]{Zinnecker:2007}
{Zinnecker}, H. \& {Yorke}, H.~W. 2007, \araa, 45, 481

\end{thebibliography}

\label{lastpage}

\end{document}